\documentclass{article}
\usepackage{amssymb}
\usepackage{epsfig}
\usepackage{amsmath}

\input{tcilatex}

\begin{document}

\title{\textbf{Periapsis and gravitomagnetic precessions of stellar orbits in Kerr and Kerr-de Sitter black hole spacetimes}}
\author{G. V. Kraniotis \footnote{kraniotis@physics.tamu.edu, MIFP-06-05.} \footnote{
Present address:Ludwig-Maximilians-Universit$\rm \ddot{a}$t, 
Arnold-Sommerfeld-Centre for Theoretical Physics,Theresienstra$\ss$e 37, 80333, M$\rm{\ddot u}$nchen, Germany,
kraniotis@theorie.physik.uni-muenchen.de}\\
%EndAName
George P. and Cynthia W. Mitchell Institute for Fundamental Physics, \\
Texas A$\&$M University,\\
 College Station, TX 77843, USA \\
}
\maketitle

\begin{abstract}
The exact solution for the motion 
of a test particle in a non-spherical polar orbit 
around a Kerr black hole is derived. 
Exact novel expressions for frame dragging (Lense-Thirring effect), 
periapsis advance and 
the orbital period are produced. The resulting formulae, are expressed 
in terms of Appell's first hypergeometric function  $F_1$, Jacobi's amplitude function, 
and Appell's $F_1$ and Gau$\ss$ hypergeometric function respectively. 
The exact expression for frame dragging is applied for the calculation 
of the Lense-Thirring effect for the orbits of S-stars in the central 
arcsecond of our Galaxy assuming that the galactic centre  is a Kerr black 
hole, for various values of the Kerr parameter including those supported 
by recent observations. 
In addition, we apply our solutions for the calculation 
of frame dragging and periapsis advance 
for stellar non-spherical polar orbits in regions of strong 
gravitational field close to the event 
horizon of the galactic black hole, e.g. for orbits in the 
central milliarcsecond of our galaxy. Such orbits are the target of 
the GRAVITY experiment. We provide examples with orbital periods in the 
range of 100min - 54 days. Detection of such stellar orbits will allow 
the possibility of measuring the relativistic effect of periapsis advance 
with high precision at the strong field realm of general relativity.
Further, an exact closed form formula for the orbital period of 
a test particle in a non-circular equatorial motion around 
a Kerr black hole is produced.
We also derive exact expressions  for the periapsis advance and the 
orbital period 
for a test particle in a non-circular equatorial motion in the Kerr 
field  in the presence of the cosmological constant in terms of 
Lauricella's fourth hypergeometric function 
$F_D$.

\end{abstract}

\bigskip

\bigskip \newpage

\section{ Introduction}

\subsection{\protect\bigskip Motivation}

Most of the celestial bodies deviate very little from spherical
symmetry, and the Schwarzschild spacetime is an appropriate
approximation for their gravitational field \cite{Karl}. However, 
for some astrophysical bodies the rotation of the mass distribution 
cannot be neglected. A more general spacetime solution of the
gravitational field equations should take this property into account.
In this respect, the Kerr solution \cite{KERR} represents, the curved 
spacetime 
geometry surrounding a rotating mass \cite{OHANIAN}. Moreover, the above 
solution 
is also important  for probing the strong gravitational 
field regime of general relativity \cite{REES}.
This is significant, since general relativity has triumphed in large-scale 
cosmology \cite{PERLMUTTER, BAHCALL, DEBERNAR,
GVKSBW}, and in predicting solar system effects on planetary orbits 
like the perihelion 
precession of Mercury with a very high precision \cite{Albert, 
Mercury} \footnote{See also about the BepiColombo science mission on Mercury http://sci.esa.int/home/bepicolombo/.}.

Due to their significance for black hole physics and 
for describing orbits around a rotating mass, exact 
solutions of Kerr geodesics with and without the contribution 
from the cosmological constant have been derived and 
studied extensively in recent years
 \cite{KraniotisKerr,KraniotisLight}.
However, few important classes of possible orbits were left out of the discussion in \cite{KraniotisKerr,KraniotisLight}. 

An important such class is the case of timelike 
$non-spherical$ $polar$ Kerr orbits.
An essential property of such non-equatorial orbits in Kerr spacetime is the 
Lense-Thirring phenomenon or frame dragging: the Kerr rotation adds longitudinal dragging 
to the relativistic orbital precession in the Schwarzschild spacetime.
Thus the orbital plane for such geodesics in Kerr  spacetime is not 
fixed but rather is dragged by the 
rotation of central mass in sharp contrast with orbital solutions in 
the static Schwarzschild case. 
It is the purpose of the present work to derive the first exact solution of 
this important class of orbits in closed analytic form. Our exact analytic 
solutions provide closed form expressions for the orbit, 
Lense-Thirring effect, periapsis advance and the corresponding orbital period. Further our 
exact results allow to  precisely 
calculate the Lense-Thirring period (frequency) for these non-zero eccentricity orbits. By construction our solutions are valid at all 
regions of gravitational strength from the strong to the  weak 
gravitational field regime. Thus they are particularly important for 
probing the strong field realm of general relativity.

Another important class studied in this work for the first time is the 
case of $equatorial\; non-circular$ timelike orbits in the presence of the cosmological 
constant {\boldmath$\Lambda$}. 
In particular, we derive the exact solutions for the periapsis advance and the 
orbital period for a test particle in a non-circular equatorial motion 
in the Kerr-de Sitter gravitational field.
The derived exact formulae in closed analytic form 
are important for the precise determination of the combined effects  
of {\boldmath$\Lambda$} and the rotation of 
central mass on planetary orbits as well as for investigating the 
effect of the cosmological constant in black hole physics e.g. accretion physics for non-zero eccentricity trajectories.

Motivated by the observational evidence that the galactic centre region 
Sgr A$^{*}$, of Milky Way,  
is the best candidate for a supermassive rotating black hole \cite{GENZEL,Porquet},
we applied our solutions for non-spherical polar Kerr orbits in two 
situations a) for test particle stellar orbits near the outer horizon of the 
galactic centre black hole (strong field regime) 
b) for calculating the frame dragging of the observed orbits of S-stars 
in the central arcsecond of our galaxy (weak field regime).

Indeed observations of the galactic nucleus at the near-infrared and 
X-rays have provided a quite compelling  evidence 
for the existence of a supermassive black hole of mass $M_{\rm BH}=(3-4)\times 
10^6M_{\odot}$ at the galactic centre, located at a distance of approximately $8$Kpc \cite{Schoedel,
GHEZa,GENZEL,Porquet,SINFONI} ($M_{\odot}$ denotes the mass of the Sun ). 

Further observations of near-infrared periodic flares have revealed 
that the central black hole is rotating so that the spacetime region near 
Sgr ${\rm A^{*}}$ is described by the Kerr geometry rather that the Schwarzschild 
static geometry, with the Kerr parameter \cite{GENZEL} 
\begin{equation}
\frac{J}{G M_{{\rm BH}}/c}=0.52\; (\pm 0.1,\pm 0.08,\pm 0.08)
\label{Galaxy}
\end{equation}
where the reported high-resolution infrared 
observations of $\rm {Sgr\; A^{*}}$
revealed `quiescent' emission and several flares. These flare events last 
for about an 1 h, and their light-curves show significant variations on 
a typical scale of 17min.
In the above equation $J$ \footnote{$J=c a$ where $a$ is the Kerr parameter. The interpretation of $ca$ as the angular momentum per unit mass was first given by Boyer and Price \cite{BOPR}. In fact, by comparing with the Lense-Thirring 
calculations \cite{LTPr} they determined the Kerr parameter to be: $a=-
\frac{2 \Omega l^2}{5 c}$, where $\Omega$ and $l$ denote the angular 
velocity and radius of the rotating sphere.}  
denotes the angular momentum of the 
black hole (The error 
estimates here the uncertainties in the period, black hole mass  
and distance to the galactic centre, respectively; $G$ is the gravitational 
constant and $c$ the velocity of light.)  This is half the maximum value for a Kerr black hole \cite{SPIN}. 
Observation of X-ray flares confirmed that the spin of the supermassive 
black hole is indeed substantial and values of the Kerr parameter as high 
as $a_{\rm Galactic}=0.9939$ have been recently reported \cite{Porquet,Aschenbach}. Let us also
mention that a hump of the locally non-rotating frames (LNRF) 
velocity profile has been found for accretion discs orbiting 
rapidly rotating Kerr black holes with spin $a>0.9953$ for 
Keplerian discs  \cite{Aschenbach} and $a>0.99979$ for marginally stable discs 
\cite{ZdenekA} \footnote{Such  humpy equatorial orbital velocity profiles 
for stable, circular orbits, occur in a small 
radial range close to but above the marginally stable equatorial 
circular geodesic of the black hole \cite{Aschenbach, ZdenekA}.}.

The material of this paper is organized as follows.
Section 2 and 3 deals with polar geodesics in a Kerr background 
while section 4 discusses equatorial geodesics in a (cosmological) 
Kerr geometry. In section 2 we derive the first exact orbital 
solution that describes the motion of a test particle in a non-spherical 
polar orbit in the Kerr gravitational field. More specifically, the 
exact expression in closed analytic form for the amount of frame dragging 
(Lense-Thirring) effect  that a test particle undergoes is given 
in Eq.(\ref{TimelikePolar}) of subsection \ref{FRAMEDRAGPOLAR}, and its 
relativistic periapsis advance is given in Eq.(\ref{PERIASTRONSHIFTP}) 
of subsection \ref{PeriastronPrecessionPT}.
In subsection \ref{PeriodTime} we derive an exact expression for the 
orbital period of a polar non-spherical 
Kerr orbit Eq.(\ref{PolarOrbitalPeriod}) 
which is important for defining the 
Lense-Thirring period Eq.(\ref{PolarFrequency}). We subsequently 
apply our results to an important system for which observational 
data is available. Namely i) in \ref{FramedragSstars}  we 
calculate the frame dragging for the observed orbits of S-stars in 
the central arcsecond of our Galaxy assuming that the galactic centre 
is a rotating Kerr black hole. We calculated the Lense-Thirring periods 
for various values of the Kerr parameter and compared the result with 
estimated ages of the S-stars from spectroscopic studies. 
ii) We calculate in section \ref{MILLITOKSODEUTEROLEPTOU} the 
frame dragging and periapsis advance for polar non-spherical Kerr 
orbits in the central milliarcsecond of our Galaxy. In section 4 
we derive the exact solution for the relativistic periastron advance, 
Eq.(\ref{LambdaPeriastronAdvance}), for the equatorial orbital 
non-circular motion of a test particle in the Kerr field in the 
presence of the cosmological constant. We also find the orbital 
period, Eq.(\ref{IsimeriniPeriodos}), for equatorial non-circular 
orbits in the asymptotically flat Kerr black hole.

The angular resolutions required for measuring stellar 
orbits in the central milliarcsecond of Milky Way  will be achievable by the General 
Relativity Analysis via \textsc{Vlt} InTerferometrY (GRAVITY) experiment 
using Very Long 
Baseline Interferometry (VLBI) techniques \cite{GRAV}. 
The detection of such stellar cusp
orbits, predicted by the theory \footnote{Stellar counts expect an 
apparent magnitude in the range $17.5\leq m_K \leq 19.5$ for stars 
within this region of 
the galactic centre.},
will allow a precise measurement of the relativistic effect of periapsis advance since the orbital periods are small and therefore will allow a test 
of the precise theory developed in this work, based on the exact 
solutions of the equations of General Relativity at the strong-field realm. 
A completely unexplored region of paramount importance for 
fundamental physics. We present examples of non-zero eccentricity 
trajectories 
with orbital periods in the range 100min-54 days. 
In a particular solution that will be presented in section \ref{MILLITOKSODEUTEROLEPTOU} with orbital 
period of 53.6 days the phenomenon of relativistic periapsis advance is 
19.8$\frac{\circ}{\rm yr}$, already a very large effect, which 
therefore should be easily measurable with a high precision, after say one year of observation. 

%%In section (\ref{PeriastronLambdaAdvance}) we derive the exact solution 
%%for the relativistic periastron advance for the equatorial orbital non-circula%%r motion of a test particle in the Kerr field in the presence of the cosmologi%%%ca%%l  constant. The derived formula is given in terms of Lauricella's fourth 
%%hypergeometric function $F_D$ of three and four variables.

In a series of appendices we present some of our formal calculations as 
well as some background on Lauricella's hypergeometric function, 
the differential equations it obeys as well as its integral representation.

Taking into account the contribution from the cosmological constant,
the generalization of the Kerr solution is described by the Kerr -de Sitter
metric element which in Boyer-Lindquist (BL) coordinates \footnote{These 
coordinates have the advantage that reduce to the Schwarzschild solution 
with a cosmological constant in the limit $a\rightarrow 0$, see \cite{Boyer}.}
is given by \cite{Zdenek,CARTER}:

\begin{eqnarray}
ds^2&=&\frac{\Delta_r}{\Xi^2 \rho^2}\left(c dt-a\sin^2
\theta d\phi\right)^2
-\frac{\rho^2}{\Delta_r}dr^2-\frac{\rho^2}{\Delta_{\theta}}d\theta^2\nonumber
\\
&-&\frac{\Delta_{\theta}\sin^2
\theta}{\Xi^2 \rho^2}\left(a c dt-(r^2+a^2)d\phi\right)^2
\end{eqnarray}
where 
\begin{eqnarray}
\Delta_r &:=&(1-\frac{\mbox{\boldmath$\Lambda$}}{3}r^2)(r^2+a^2)-\frac{2 G M_{\rm BH} }{c^2}r \nonumber \\
\Delta_{\theta}&:=&1+\frac{a^2 \mbox{\boldmath$\Lambda$}}{3}\cos^2 \theta \nonumber \\
\Xi&:=&1+\frac{a^2 \mbox{\boldmath$\Lambda$}}{3},\;\;\rho^2:=r^2+a^2  \cos^2 \theta \nonumber \\
 \end{eqnarray}
and {\boldmath$\Lambda$} denotes the cosmological constant, $a$ is the 
Kerr (rotation) parameter and $M_{\rm BH}$ is the mass of the black hole.

The timelike geodesic equations in the presence of the cosmological constant have been 
derived in \cite{KraniotisKerr} by solving the Hamilton-Jacobi partial 
differential equation with the method of separation of variables:

\begin{eqnarray}
\int^r \frac{dr}{\sqrt{R}}=\pm\int^{\theta} \frac{d\theta}{\sqrt{\Theta}}
\nonumber \\
\rho^2 \frac{d\phi}{ds}=-\frac{\Xi^2}{\Delta_{\theta}\sin^2\theta}
\left(aE\sin^2\theta-L\right)+\frac{a\Xi^2}{\Delta_r}\left[(r^2+a^2)E-aL\right]
\nonumber \\
c\rho^2 \frac{dt}{ds}=\frac{\Xi^2(r^2+a^2)\left[(r^2+a^2)E-aL\right]}{\Delta_r}-\frac{a\Xi^2 (aE \sin^2\theta-L)}{\Delta_{\theta}} \nonumber \\
\rho^2\frac{dr}{ds}=\pm \sqrt{R} \nonumber \\
\rho^2\frac{d\theta}{ds}=\pm \sqrt{\Theta} 
\label{LambdaGeo}
\end{eqnarray}
where 
\begin{eqnarray}
R&:=&\Xi^2 \left[(r^2+a^2)E-aL\right]^2-\Delta_r\left(r^2+Q+
\Xi^2(L-aE)^2\right) \nonumber \\
\Theta&:=&\left[Q+(L-aE)^2 \Xi^2-a^2 \cos^2\theta\right]\Delta_{\theta}-\Xi^2\frac{(aE \sin^2\theta-L)^2}{\sin^2\theta}
\label{LARTH}
\end{eqnarray}

The constants of integration $E,L$ are
 associated with the isometries of the Kerr metric. Carter's constant of 
integration is denoted by $Q$.  

The geodesic differential equations (\ref{LambdaGeo}), (\ref{LARTH}) 
reduce to the ones found by Carter \cite{carter2} 
when we set the cosmological constant to zero.

\section{Exact solution of non-spherical polar timelike Kerr geodesics}
\label{MSPHAIRAPOL}
\subsubsection{Frame dragging effect for polar non-spherical bound orbits}
\label{FRAMEDRAGPOLAR}

In this section we shall derive the exact solution for the important class 
of non-spherical polar orbits of a test particle in the gravitational field 
of a Kerr black hole.
We shall first
determine the change in the azimuthal angle $\phi$ after a 
complete radial oscillation for the case of 
non-spherical polar timelike geodesics, assuming a vanishing 
cosmological constant. This will allow us to calculate the resulting 
frame-dragging (Lense-Thirring effect).
Thus we generalize our earlier results in which the exact solutions 
of the spherical polar timelike geodesics have been derived and 
applied for the determination of frame dragging (Lense-Thirring)
 effect for stellar 
orbirs close to the galactic centre, and for satellite orbits around the 
Earth \cite{KraniotisKerr,KraniotisLight}. It has been shown \cite{StoTsou} that a necessary condition  for an orbit to be $polar$ (meaning to intersect the symmetry axis of the 
Kerr gravitational field)  is the vanishing of the parameter $L$, i.e . $L=0$
\footnote{Such orbits are also characterized by  a non-vanishing value for 
the hidden first integral represented by Carter's constant $Q$.}.
The relevant differential equation for the calculation of frame dragging is 
\begin{equation}
\frac{d\phi}{dr}=\frac{2 \frac{G M_{\rm BH}}{c^2}a E r}{\Delta \sqrt{R}}
\label{FRAMEDRAG}
\end{equation}
where the quartic polynomial $R$, for non-spherical 
polar Kerr orbits, is given by (\ref{LARTH}) with 
{\boldmath$\Lambda$}$=0,L=0$. Also $\Delta:=r^2+a^2-\frac{2 G M_{\rm BH} r}{c^2}$. 
%%\begin{equation}
%%R=(r^2+a^2)^2 E^2-\Delta(r^2+a^2 E^2+Q)
%%\label{PolarQuartic}
%%\end{equation}
As in \cite{KraniotisLight} we will use partial fractions for 
integrating equation (\ref{FRAMEDRAG}) from periapsis distance 
$r_P$ to apoapsis  distance $r_A$.
Thus we have 
\begin{eqnarray}
\frac{d\phi}{dr}
&=&\frac{A_+^P}{(r-r_+)\sqrt{R}}+\frac{A_-^P}{(r-r_-)\sqrt{R}} \nonumber \\
&=&\frac{A_+^P}{(r-r_+)\sqrt{(E^2-1)(r-\alpha)(r-\beta)(r-\gamma)(r-\delta)}
}\nonumber \\
&+&
\frac{A_-^P}{(r-r_-)\sqrt{(E^2-1)(r-\alpha)(r-\beta)(r-\gamma)(r-\delta)}
} \nonumber \\
\end{eqnarray}
We denote the real roots of the quartic polynomial by $\alpha,\beta,\gamma,\delta$, $\alpha>\beta>\gamma>\delta$ and 
 organize all roots in ascending 
order of magnitude as follows 
\begin{equation}
\alpha_{\rho}>\alpha_{\sigma}>\alpha_{\nu}>\alpha_i
\end{equation}
where $\alpha_{\rho}=\alpha_{\mu},\alpha_{\sigma}=\alpha_{\mu+1},
\alpha_{\nu}=\alpha_{\mu+2}, \alpha_i=\alpha_{\mu-i},i=1,2,3$ and we have 
that $\alpha_{\mu-1}\geq \alpha_{\mu-2}>\alpha_{\mu-3}$.
By applying the transformation  \footnote{We have the correspondence: $\alpha_{\mu}=
\alpha,\alpha_{\mu+1}=\beta,\alpha_{\mu+2}=\gamma,\alpha_{\mu-1}=
\alpha_{\mu-2}=r_{\pm}^{\prime},\alpha_{\mu-3}=\delta$.}
\begin{equation}
z=\frac{1}{\omega}\frac{r^{\prime}-\alpha_{\mu+1}}{r^{\prime}-\alpha_{\mu+2}}
\label{SymaMetasxi}
\end{equation}
where 
\begin{equation}
\omega=\frac{\alpha_{\mu}-\alpha_{\mu+1}}{\alpha_{\mu}-\alpha_{\mu+2}}
\end{equation}
and after a dimensionless variable $r^{\prime}$ through  $r=r^{\prime}\frac{GM_{\rm BH}}{c^2}$ has been introduced,
we can bring the radial integral which determines the amount of frame dragging 
that a non-spherical polar Kerr orbit undergoes 
in the general theory of relativity (GTR)
\begin{equation}
\Delta\phi^{\rm GTR}=2 \int_{r_P}^{r_A}\frac{2 \frac{G M_{\rm BH}}{c^2}a E r}{\Delta \sqrt{R}}dr
\end{equation}
into the familiar integral representation of Appell's first 
hypergeometric function of two variables $F_1$.

Indeed we obtain 
\begin{eqnarray}
\Delta\phi^{\rm {GTR}}&=&2\Biggl[-\frac{\omega^{3/2}A_+^P}{H_+}
F_1\left(\frac{3}{2},1,\frac{1}{2},2,\kappa^2_+,\mu^2\right)\frac{\Gamma\left(\frac{3}{2}\right)\Gamma\left(\frac{1}{2}\right)}{\Gamma(2)} \nonumber \\
&+&
\frac{\omega^{1/2}A_+^P}{H_+}F_1\left(\frac{1}{2},1,\frac{1}{2},1,\kappa^2_+,\mu^2\right)\frac{\Gamma^2\left(\frac{1}{2}\right)}{\Gamma(1)}+ \nonumber \\
&-&\frac{\omega^{3/2}A_-^P}{H_-}
F_1\left(\frac{3}{2},1,\frac{1}{2},2,\kappa^2_-,\mu^2\right)\frac{\Gamma\left(\frac{3}{2}\right)\Gamma\left(\frac{1}{2}\right)}{\Gamma(2)} \nonumber \\
&+&
\frac{\omega^{1/2}A_-^P}{H_-}F_1\left(\frac{1}{2},1,\frac{1}{2},1,\kappa^2_-,\mu^2\right)\frac{\Gamma^2\left(\frac{1}{2}\right)}{\Gamma(1)}\Biggr] \nonumber \\
\label{TimelikePolar}
\end{eqnarray}
The function
$F_1(\alpha,\beta,\beta^{\prime},
\gamma,x,y)$ is the first of the four Appell's hypergeometric functions of 
two variables $x,y$ \cite{APPELL} \footnote{The expression $(\lambda,\kappa)=
\lambda(\lambda+1)\cdots(\lambda+\kappa-1)$, and the symbol $(\lambda,0)$ 
represents $1$.},
\begin{equation}
F_1(\alpha,\beta,\beta^{\prime},
\gamma,x,y)=\sum_{m=0}^{\infty}\sum_{n=0}^{\infty}\frac{(\alpha,m+n)(\beta,m)(\beta^{\prime},n)}
{(\gamma,m+n)(1,m)(1,n)}x^m y^n
\end{equation}
which admits the following integral representation
\begin{equation}
\int_0^1 u^{\alpha-1}(1-u)^{\gamma-\alpha-1}(1-u x)^{-\beta}
(1-uy)^{-\beta^{\prime}}du=\frac{\Gamma(\alpha)\Gamma(\gamma-\alpha)}{
\Gamma(\gamma)}F_1 (\alpha,\beta,\beta^{\prime},\gamma,x,y)
\end{equation}
The double series converges when $|x|<1$ and $|y|<1$. The above Euler integral 
representation is valid for ${\rm Re}(\alpha)>0$ and ${\rm Re}(\gamma-\alpha)>0$. Also $\Gamma(p)=\int_0^{\infty}x^{p-1}e^{-x}dx$ denotes the gamma function.
In addition
\begin{eqnarray}
H_{\pm}:&=&\sqrt{(1-E^2)}(\alpha_{\mu+1}-\alpha_{\mu-1})\sqrt{\alpha_{\mu}-
\alpha_{\mu+1}}\sqrt{\alpha_{\mu+1}-\alpha_{\mu-3}} \nonumber \\
&=&\sqrt{(1-E^2)}(\beta-r_{\pm}^{\prime})\sqrt{\alpha-\beta}\sqrt{\beta-\delta} 
\end{eqnarray}
and 
\begin{equation}
A_{\pm}^P=-(\pm)\frac{2 a E  r^{\prime}_{\pm}}{r_{-}^{\prime} - r_+^{\prime}}
\label{SYNTEADIA}
\end{equation}
The radii of the event horizon and the inner or Cauchy horizon, 
$r_{+},r_-$ respectively of the black hole are located at
\begin{equation}
r_{\pm}=\frac{GM_{\rm BH}}{c^2}\pm \sqrt{\left(\frac{GM_{\rm BH}}{c^2}\right)^2-a^2}
\end{equation}
In equation (\ref{SYNTEADIA}) $r^{\prime}_{\pm}:=\frac{r_{\pm}}{\frac{GM_{\rm BH}}{c^2}}$, $a$ are the 
dimensionless horizon radii and spin of the black hole respectively.

The moduli, introduced in (\ref{TimelikePolar}), are given in terms of the roots of the quartic and the horizon radii of the rotating black hole by the expressions 
\begin{eqnarray}
\kappa_{\pm}^2&=&\frac{\alpha_{\mu}-\alpha_{\mu+1}}{\alpha_{\mu}-\alpha_{\mu+2}}
\frac{\alpha_{\mu-1}-\alpha_{\mu+2}}{\alpha_{\mu-1}-\alpha_{\mu+1}}=
\frac{\alpha-\beta}{\alpha-\gamma}\frac{r_{\pm}^{\prime}-\gamma}{r_{\pm}^{\prime}-\beta}, \nonumber \\
\mu^2&=&\frac{\alpha_{\mu}-\alpha_{\mu+1}}{\alpha_{\mu}-\alpha_{\mu+2}} \frac{\alpha_{\mu-3}-\alpha_{\mu+2}}{\alpha_{\mu-3}-\alpha_{\mu+1}}=
\frac{\alpha-\beta}{\alpha-\gamma}\frac{\delta-\gamma}{\delta-\beta}, \nonumber \\
\end{eqnarray}
Equation (\ref{TimelikePolar}) represents the $first$ 
exact formula  of frame dragging 
(Lense-Thirring effect) that a test particle in a non-spherical polar 
bound orbit  around a Kerr black hole undergoes. It represents a generalization of our previous results 
for test particles and photons in spherical polar orbits \cite{KraniotisKerr,
KraniotisLight}.

\subsubsection{Periapsis advance for non-spherical polar orbits}
\label{PeriastronPrecessionPT}
 
In this subsection we shall investigate the periastron advance for a 
non-spherical bound Kerr polar orbit, assuming a vanishing cosmological 
constant. We shall first obtain the exact solution for the orbit and 
then derive an exact  closed form formula for the periapsis precession.
The relevant differential equation is
\begin{equation}
\int^r\frac{dr}{\sqrt{R}}=\pm \int^{\theta} \frac{d\theta}{\sqrt{\Theta}}
\label{PAPTM}
\end{equation}
where in (\ref{PAPTM}), $\Theta$ and $R$ are quantities defined, 
respectively, in (\ref{LARTH}) with {\boldmath$\Lambda$}$=0,L=0$.
%%\begin{equation}
%%\Theta=Q-\left[a^2(1-E^2)\right]\cos^2 \theta
%%\end{equation}
Now applying  the transformation (\ref{SymaMetasxi}) on the left hand side 
of equation (\ref{PAPTM}) we get
\begin{eqnarray}
\int\frac{dr}{\sqrt{R}}&=&\frac{1}{\frac{GM_{\rm BH}}{c^2}}\int\frac{dr^{\prime}}{\sqrt{(1-E^2)(-)
(r^{\prime}-\alpha)(r^{\prime}-\beta)(r^{\prime}-\gamma)(r^{\prime}-\delta)}} \nonumber \\
&=&\frac{1}{\frac{GM_{\rm BH}}{c^2}}\int\frac{dz \sqrt{\omega}}{\sqrt{(1-E^2)}
\sqrt{(\alpha-\beta)(\beta-\delta)}\sqrt{z(1-z)(1-\kappa^2 z)}}
\end{eqnarray}
where \footnote{The roots of the quartic are organized as: 
$\alpha_{\mu}>\alpha_{\mu+1}>\alpha_{\mu+2}>\alpha_{\mu-3}$, and 
we have the correspondence $\alpha=\alpha_{\mu}, \beta=\alpha_{\mu+1},\gamma=\alpha_{\mu+2}, 
\delta=\alpha_{\mu-3}$.}
\begin{eqnarray}
\kappa^2&=&\frac{\alpha-\beta}{\alpha-\gamma}\frac{\delta-\gamma}{\delta-\beta
}=\omega \frac{\delta-\gamma}{\delta-\beta
}
\end{eqnarray}
%and 
%\begin{equation}
%\omega=\frac{\alpha-\beta}{\alpha-\gamma}
%\end{equation}

By setting, $z=x^2$, we obtain the equation \footnote{The roots of the 
quartic and the angular integration are understood to be dimensionless in 
this equation.}
\begin{equation}
\int\frac{dx}{\sqrt{(1-x^2)(1-\kappa^2 x^2)}}=
\frac{\sqrt{1-E^2}\sqrt{\alpha-\beta}\sqrt{\beta-\delta}}{2 \sqrt{\omega}}
\int\frac{d\theta}{\sqrt{\Theta}}
\end{equation}
Using the idea of $invertion$ for the orbital 
elliptic integral on the left hand side 
we  
obtain
\begin{equation}
x={\rm sn}\left(\frac{\sqrt{1-E^2}\sqrt{(\alpha-\gamma)(\beta-\delta)}}{2}\int
\frac{d\theta}{\sqrt{\Theta}},\kappa^2\right)
\end{equation}
In terms of the original variables we derive the 
equation
\begin{equation}
r=\frac{\beta-\gamma \frac{\alpha-\beta}{\alpha-\gamma}
{\rm sn^2}\left(\frac{\sqrt{1-E^2}\sqrt{(\alpha-\gamma)(\beta-\delta)}}{2}
\int\frac{d\theta}{\sqrt{\Theta}},\kappa^2\right)}{
1-\frac{\alpha-\beta}{\alpha-\gamma}
{\rm sn^2}\left(\frac{\sqrt{1-E^2}\sqrt{(\alpha-\gamma)(\beta-\delta)}}{2}
\int\frac{d\theta}{\sqrt{\Theta}},\kappa^2\right)}\frac{GM_{\rm BH}}{c^2}
\label{EXACTPOLARORBIT}
\end{equation}
Equation (\ref{EXACTPOLARORBIT}) represents the first exact solution 
that describes the motion of a test particle in a polar non-spherical 
bound orbit in the Kerr field in terms of Jacobi's sinus 
amplitudinous elliptic function.
The function ${\rm sn^2}(y,\kappa^2)$ has period 
$2K(\kappa^2)=\pi F\left(\frac{1}{2},\frac{1}{2},1,\kappa^2\right)$ which is 
also the period of $r$ \footnote{$K(\kappa^2)$ denotes the complete elliptic 
integral of the first kind.}. 
The Gau$\ss$ hypergeometric function $F(\alpha_1,\beta_1,\gamma_1,x)$ is equal to  $1+\frac{\alpha_1.\beta_1}{1.\gamma_1}x+\frac{\alpha_1(\alpha_1+1)\beta_1(\beta_1+1)}{1.2.\gamma_1(\gamma_1+1)}x^2+\cdots$ \cite{KUMMER}.
This means that after one complete 
revolution the angular integration has to satisfy the equation
\begin{equation}
-\int\frac{d\theta}{\sqrt{\Theta}}=\frac{1}{\sqrt{Q}}\int\frac{\partial \Psi}{\sqrt{1-\kappa^{\prime 2} \sin^2\Psi}}=
\frac{4}{\sqrt{1-E^2}}\frac{1}{\sqrt{\alpha-\gamma}}\frac{1}{
\sqrt{\beta-\delta}}\frac{\pi}{2}F\left(\frac{1}{2},\frac{1}{2},1,\kappa^2\right)
\label{GwniakiOloklirosi}
\end{equation}
where the latitude  variable
\begin{equation} 
\Psi:=\pi/2-\theta
\end{equation} 
has been introduced.
Equation (\ref{GwniakiOloklirosi}) can be rewritten as
\begin{eqnarray}
& &\int\frac{dx^{\prime}}{\sqrt{(1-x^{\prime 2})( 1-\kappa^{\prime 2}x^{\prime 2})}} \nonumber \\
&=&\sqrt{Q}\frac{4}{\sqrt{1-E^2}}\frac{1}{\sqrt{\alpha-\gamma}}\frac{1}{
\sqrt{\beta-\delta}}\frac{\pi}{2}F\left(\frac{1}{2},\frac{1}{2},1,\kappa^2\right)
\label{AngularPeriapsis}
\end{eqnarray}
and 
\begin{equation}
\kappa^{\prime 2}:=\frac{a^2 (1-E^2)}{Q}
\label{KappaPrimeSq}
\end{equation}
Also $x^{\prime 2}=\sin^2 \Psi=\cos^2 \theta$.
Equation (\ref{AngularPeriapsis}) determines the exact amount 
that the angular integration satisfies  after a 
complete radial oscillation  for a bound non-spherical polar orbit.

Now using the latitude  variable $\Psi$ the change in latitude  
after a complete radial oscillation leads to the following
exact expression for the periastron advance for a test particle in a 
non-spherical polar Kerr orbit, assuming a vanishing cosmological constant,
\begin{eqnarray}
\Delta \Psi^{\rm GTR}&=&\Delta\Psi-2\pi \nonumber \\
&=&am\left(\sqrt{Q}\frac{4}{\sqrt{1-E^2}}\frac{1}{\sqrt{\alpha-\gamma}}\frac{1}{
\sqrt{\beta-\delta}}\frac{\pi}{2}F\left(\frac{1}{2},\frac{1}{2},1,\kappa^2
\right),\frac{a^2 (1-E^2)}{Q}\right)-2 \pi \nonumber \\
\label{PERIASTRONSHIFTP}
\end{eqnarray}
where the Abel-Jacobi's amplitude $am(\mu,\kappa^{\prime 2})$ is the function that inverts the 
elliptic integral  
\begin{equation}
\int_0^{\Psi}\frac{\partial \Psi}{\sqrt{1-\kappa^{\prime 2} \sin^2\Psi}}=\mu
\end{equation}
in other words $\Psi=am(\mu,\kappa^{\prime 2})$.
For those astrophysical applications for which the modulus $\kappa^{\prime 2}=\frac{a^2 (1-E^2)}{Q}
\rightarrow 0$, $\Delta \Psi^{\rm GTR}=\sqrt{Q}\frac{4}{\sqrt{1-E^2}}\frac{1}{\sqrt{\alpha-\gamma}}\frac{1}{
\sqrt{\beta-\delta}}\frac{\pi}{2}F\left(\frac{1}{2},\frac{1}{2},1,\kappa^2,
\right)-2\pi$ \footnote{Alternatively this can be seen from the fact that 
$am(u,\kappa^2)={\rm ArcSin}({\rm sn}(u,\kappa^2))$ and ${\rm sn}(u,0)=\sin(u)$.}.
If we assume an initial value $\Psi_0$ for the latitude variable $\Psi$ different from zero then 
\begin{eqnarray}
\Psi&=&am\Biggl(\sqrt{Q}\frac{4}{\sqrt{1-E^2}}\frac{1}{\sqrt{\alpha-\gamma}}\frac{1}{\sqrt{\beta-\delta}}\frac{\pi}{2}F\left(\frac{1}{2},\frac{1}{2},1,\kappa^2
\right) \nonumber \\
&+ & \sin\Psi_0\;F_1\left(
\frac{1}{2},\frac{1}{2},\frac{1}{2},\frac{3}{2},
\sin^2 \Psi_0,\kappa^{\prime 2} \sin^2 \Psi_0\right)
,\frac{a^2 (1-E^2)}{Q}\Biggr) \nonumber \\
\label{PERIASTRONSHIFTP1}
\end{eqnarray}
and $\Delta\Psi^{\rm GTR}=\delta \Psi-2 \pi=(\Psi-\Psi_0)-2\pi $ \footnote{
$\int_0^{\Psi_0}\frac{d\Psi}{\sqrt{1-\kappa^{\prime 2}\sin^2(\Psi)}}=
\sin\Psi_0 F_1\left(\frac{1}{2},\frac{1}{2},\frac{1}{2},\frac{3}{2},
\sin^2 \psi_0,\kappa^{\prime 2}\sin^{2} \Psi_0\right)$ see Appendix 
\ref{AngularOloklirosi} for the exact derivation of this angular integration.}.

The orbital solution of non-polar non-spherical geodesics is discussed in 
Appendix \ref{nosphnp}.

\subsection{Calculation of the orbital period in non-spherical polar 
Kerr geodesics}
\label{PeriodTime}
In this subsection we shall perform an exact calculation of the period 
${\rm P}$ for the non-spherical polar orbit of a test particle in Kerr gravitational 
field. The relevant differential equation is 
\begin{equation}
\frac{c dt}{dr}=\frac{r^2+a^2}{\Delta \sqrt{R}}E(r^2+a^2)-\frac{a^2 E \sin^2\theta}{\sqrt{R}}
\end{equation}
and we integrate from periapsis to apoapsis and back to periapsis.

The physical quantity $\rm {P}$, is an important relativistic parameter, since orbits with small period 
and close to the event horizon exhibit large relativistic effects which 
are easily measurable. 
The resulting expression has an elegant analytic form in terms of 
generalized hypergeometric functions of Appell and Gau\ss's 
hypergeometric function. We obtain
\begin{eqnarray}
ct\equiv c{\rm P}&=&\frac{E \beta^2 2 \frac{GM_{\rm BH}}{c^2}}{\sqrt{1-E^2}\sqrt{\alpha-\gamma}\sqrt{\beta-\delta}}
\Biggr[\frac{\Gamma^2 (1/2)}{\Gamma(1)}F_1\left(\frac{1}{2},2,\frac{1}{2},1,\omega,\kappa^2\right) \nonumber \\
&-&\frac{2\omega \gamma}{\beta}\frac{\Gamma(3/2)\Gamma(1/2)}{\Gamma(2)}
F_1\left(\frac{3}{2},2,\frac{1}{2},2,\omega,\kappa^2\right)+\frac{\gamma^2 \omega^2}{\beta^2}\frac{\Gamma(5/2)\Gamma(1/2)}{\Gamma(3)}F_1\left(\frac{5}{2},2,\frac{1}{2},3,\omega,\kappa^2\right)\Biggr]  \nonumber \\
&+& Ea^2  2 \frac{GM_{\rm BH}}{c^2}\sqrt{\frac{\omega}{1-E^2}}\frac{1}{\sqrt{(\alpha-\beta)(\beta-\delta)}}F(1/2,1/2,1,\kappa^2) 
\frac{\Gamma^2(1/2)}{\Gamma(1)} \nonumber \\
&+&\frac{4 E G M_{\rm{BH}}}{c^2} \sqrt{\frac{\omega}{1-E^2}}\frac{\beta}{\sqrt{(\alpha-\beta)(\beta-\delta)}}
\Biggl[\frac{\Gamma^2 (\frac{1}{2})}{\Gamma(1)}F_1\left(\frac{1}{2},1,\frac{1}{2},1,\omega,\kappa^2\right) \nonumber \\
&-&
\frac{\omega \gamma}{\beta} F_1\left(\frac{3}{2},1,\frac{1}{2},2,\omega,\kappa^2\right)
\frac{\Gamma\left(\frac{3}{2}\right)\Gamma\left(\frac{1}{2}\right)}
{\Gamma(2)}\Biggr] \nonumber \\
&+&\frac{4 EG 2 M_{\rm{BH}}}{c^2}\sqrt{\frac{\omega}{1-E^2}}\frac{1}{\sqrt{(\alpha-\beta)(\beta-\delta)}}F(1/2,1/2,1,\kappa^2) 
\frac{\Gamma^2(1/2)}{\Gamma(1)} \nonumber \\
&+&\frac{4 E 2 G M_{\rm{BH}}}{c^2} \Biggl[-\frac{\omega^{3/2}A_+}{H_+}
F_1\left(\frac{3}{2},1,\frac{1}{2},2,\kappa^2_+,\mu^2\right)\frac{\Gamma\left(\frac{3}{2}\right)\Gamma\left(\frac{1}{2}\right)}{\Gamma(2)} \nonumber \\
&+&
\frac{\omega^{1/2}A_+}{H_+}F_1\left(\frac{1}{2},1,\frac{1}{2},1,\kappa^2_+,\mu^2\right)\frac{\Gamma^2\left(\frac{1}{2}\right)}{\Gamma(1)}+ \nonumber \\
&-&\frac{\omega^{3/2}A_-}{H_-}
F_1\left(\frac{3}{2},1,\frac{1}{2},2,\kappa^2_-,\mu^2\right)\frac{\Gamma\left(\frac{3}{2}\right)\Gamma\left(\frac{1}{2}\right)}{\Gamma(2)} \nonumber \\
&+&
\frac{\omega^{1/2}A_-}{H_-}F_1\left(\frac{1}{2},1,\frac{1}{2},1,\kappa^2_-,\mu^2\right)\frac{\Gamma^2\left(\frac{1}{2}\right)}{\Gamma(1)}\Biggr] \nonumber \\
&+&\frac{- a^2 E \frac{G M_{\rm{BH}}}{c^2}}{\sqrt{Q}}\Biggl[\sin(\varphi)F_1\Biggl(\frac{1}{2},\frac{1}{2},\frac{1}{2},\frac{3}{2},
\sin^2\varphi,\kappa^{2\prime}\sin^2\varphi\Biggr) \nonumber \\
&+&\Biggl\{-\sin(\varphi)F_1\Biggl(\frac{1}{2},\frac{1}{2},\frac{1}{2},\frac{3}{2},
\sin^2\varphi,\kappa^{2\prime}\sin^2\varphi\Biggr) \nonumber \\
&+&\sin(\varphi)F_1\Biggl(\frac{1}{2},\frac{1}{2},-\frac{1}{2},\frac{3}{2},
\sin^2\varphi,\kappa^{2\prime}\sin^2\varphi\Biggr)\Biggr\} \times\frac{1}{\kappa^{2\prime}}\Biggr]  \nonumber \\
\label{PolarOrbitalPeriod}
\end{eqnarray}
where $\kappa^{\prime2}$ is given by (\ref{KappaPrimeSq})
and 
\begin{equation}
\varphi=am\left(\sqrt{Q}\frac{4}{\sqrt{1-E^2}}\frac{1}{\sqrt{\alpha-\gamma}}\frac{1}{
\sqrt{\beta-\delta}}\frac{\pi}{2}F\left(\frac{1}{2},\frac{1}{2},1,\kappa^2
\right),\kappa^{\prime2}\right)
\end{equation}
Also 
\begin{equation}
A_+:=-\frac{a^2-2 r^{\prime}_+}{r^{\prime}_{-}-r^{\prime}_+},\;A_-:=-\frac{-a^2+2 r^{\prime}_{-}}{r^{\prime}_{-}-r^{\prime}_+}
\label{Syntelestes1}
\end{equation}
and
$H_{\pm}:=\sqrt{(1-E^2)(\alpha-\beta)}(r^{\prime}_{\pm}-\beta)
\sqrt{\beta-\delta}$.
The moduli (variables) of the hypergeometric function of Appell are 
given by
\begin{eqnarray}
\mu^2=\kappa^2=\frac{\alpha-\beta}{\alpha-\gamma}\frac{\delta-\gamma}{\delta-\beta} \nonumber \\
\kappa^2_{\pm}=\frac{\alpha-\beta}{\alpha-\gamma}\frac{r^{\prime}_{\pm}-\gamma}{r^{\prime}_{\pm}-\beta} \nonumber \\
\label{ModuliAppell}
\end{eqnarray}

The Lense-Thirring Period (LTP), is defined in terms of Lense-Thirring's 
precession equation (\ref{TimelikePolar}) and the orbital period ${\rm P}$, equation 
(\ref{PolarOrbitalPeriod}),
as follows
\begin{equation}
{\rm LTP}:=\frac{2\pi {\rm P}}{\Delta\phi^{\rm GTR}}
\label{PolarFrequency}
\end{equation}
therefore its expression in closed analytic form is given entirely in terms of 
the generalized hypergeometric function of Appell $F_1$ and Gau$\ss$'s 
hypergeometric function $F$. It constitutes the first exact expression for
this important parameter 
for non-spherical polar Kerr orbits.

\subsection{Frame dragging of observed stellar orbits of S-stars}
\label{FramedragSstars}

We wish now to apply our exact formula, equation (\ref{TimelikePolar}) for 
the calculation of the 
gravitomagnetic effect (Lense-Thirring effect) for the orbits of 
S-stars in the central arcsecond of Milky Way assuming that the 
galactic centre region Sgr A$^{*}$ is a rotating Kerr black hole.

We note that   the careful study and observation of three-dimensional stellar 
orbits of S-stars at the central 
arcsecond of Milky Way provided the first 
strong evidence for the existence of a supermassive 
black hole at the galactic centre region  Sgr ${\rm A^{*}}$ \cite{Schoedel,
GHEZa,GENZEL,Porquet,SINFONI}. 
Indeed, fitting of Keplerian orbits with experimental data 
revealed that the focus of the ellipse is essentially located 
at the radio position of Sgr ${\rm A^{*}}$ \footnote{The J2000 equatorial 
coordinates of Sgr  ${\rm A^{*}}$ measured at 1996.25 were found 
to be at: R.A.=$17^{{\rm h}}45^{\rm {min}}40^{\rm s}.0409$, 
decl.=\;-$29^{\circ}00^{'}28^{''}.118$ \cite{MJREIDHAFT}.}. As a matter of fact, 
Sch${\rm {\ddot o}}$del et al \cite{Schoedel} and Ghez et al  
\cite{GHEZa} have shown that the focus 
of the orbit of the S-star S2 is located within a few 
milli-arcseconds of the nominal 
radio position of   Sgr  ${\rm A^{*}}$ \footnote{
As is explained in \cite{GEOMRO}, the observed stellar orbits can lead to a more precise determination of the distance from the Sun to the centre of the galaxy. It is well known 
that the distance $R_0$ from the Sun to the galactic centre not only is important for determining 
the structure of Milky Way but also holds an important role in establishing 
the extra galactic distance scale. It also affects estimates of the dark matter in the Local Group by affecting the Andromeda infall speed \cite{REID}.} .
The observed S-stars are B main-sequence stars as can be inferred from their 
H I Br$\gamma$ absorption \footnote{The wavelength 
of the Brackett-$\gamma$ line is $\lambda({\rm Br}\gamma-{\rm line})=2165.5$nm.}. In particular S2 whose complete orbit 
has been observed appears to be a 15-20 $M_{\odot}$ main sequence O8-B0 
 dwarf star with an estimated age $t_{S2}^{age}<10$Myr \cite{GHEZa}. 
A remark is in action: since $m_S\ll M_{\rm {BH}}$ i.e. the masses of S-stars 
are very small compared to the central black hole mass the effect of 
their gravitational fields on the black hole's gravitational field is negligible. Therefore the approximation that they move on a timelike geodesic of the 
four dimensional world is entirely justifiable and consistent . They can 
be treated as test particles.

%\label{EINSTEINPolarNStime2}
%\end{table}
 \begin{table}
\begin{center}
\begin{tabular}{|c|c|c|c|c|c||}\hline\hline
$\bf {star}$ & $Q$ & $E$ & $\Delta\phi^{\rm GTR}$ & LTP(yr) & P (yr)\\ \hline
S1 & 71705.3106 & 0.999993919 &0.0702036 $\frac{\rm arcs}{\rm revolution}$ & $1.73\times 10^9$ & 93.9  \\
S2 &  5693.30424 & 0.999979485 & 3.14297  $\frac{\rm arcs}{\rm revolution}$ & $6.25\times 10^6$ & 15.15\\
S8 &  9238.88965 & 0.999992385 & 1.51944  $\frac{\rm arcs}{\rm revolution}$ &
$5.71\times 10^7$ & 67.0\\
S12 & 10641.9649 & 0.999991213 & 1.22889  $\frac{\rm arcs}{\rm revolution}$ &$5.70\times 10^7$ & 54.05\\
S13 & 36897.1620 & 0.999988560 & 0.190214 $\frac{\rm arcs}{\rm revolution}$
& $2.48\times 10^8$ & 36.4 \\
S14 & 5321.06355 & 0.999988863 & 3.47912 $\frac{\rm arcs}{\rm revolution}$ &$1.41\times 10^7$ &37.9\\
\hline \hline
\end{tabular}
\end{center}
\caption{Predictions for Lense-Thirring effect $\Delta\phi^{\rm {GTR}}$, 
LTP and 
orbital period $\rm P$ for the observed orbits of 
S-stars around the galactic centre rotating supermassive 
black hole. The Kerr 
parameter was chosen to be $a_{\rm Galactic}=0.52\frac{GM_{\rm BH}}{c^2}$.The values 
of Carter's constant $Q$ are in units of $\frac{G^2 M_{\rm BH}^2}{c^4}$. }
\label{EINNPolarNStime1}
\end{table}

\begin{table}
\begin{center}
\begin{tabular}{|c|c|c|c|c|c||}\hline\hline
$\bf {star}$ &  $Q$ & $E$ & $\Delta\phi^{\rm GTR}$ & LTP(yr) & ${\rm P}$(yr)\\ \hline
S1 & 71705.3106 & 0.999993919 &0.134183 $\frac{\rm arcs}{\rm revolution}$ 
& $9.07\times 10^8$ & 93.9\\
S2 & 5693.30424 & 0.999979485 & 6.0073 $\frac{\rm arcs}{\rm revolution}$
&$3.27\times 10^6$ &15.15 \\
S8 & 9238.88965 & 0.999992385 & 2.90417 $\frac{\rm arcs}{\rm revolution}$ &$2.99\times 10^7$ & 67.0\\
S12 & 10641.9649 & 0.999991213 & 2.34883  $\frac{\rm arcs}{\rm revolution}$ &$2.98\times 10^7$ & 54.05\\
S13 & 36897.1620 & 0.999988560 & 0.363565 $\frac{\rm arcs}{\rm revolution}$
& $1.3\times 10^8$ & 36.4 \\
S14 & 5321.06355 & 0.999988863 & 6.64981 $\frac{\rm arcs}{\rm revolution}$&$7.39\times 10^6$ & 37.9 \\
\hline \hline
\end{tabular}
\end{center}
\caption{Predictions for Lense-Thirring effect $\Delta\phi^{\rm {GTR}}$, 
LTP and orbital period $\rm P$  for the observed orbits of 
  S-stars around the rotating galactic centre black hole. The Kerr parameter 
was chosen to be $a_{\rm Galactic}=0.9939\frac{GM_{\rm BH}}{c^2}$. The values 
of Carter's constant $Q$ are in units of $\frac{G^2 M_{\rm BH}^2}{c^4}$.}
\label{EINNPolarNStime2}
\end{table}

\begin{table}
\begin{center}
\begin{tabular}{|c|c|c|c|c|c||}\hline\hline
$\bf {star}$ &  $Q$ & $E$ & $\Delta\phi^{\rm GTR}$ & LTP(yr) & ${\rm P}$(yr)\\ \hline
S1 & 71705.3106 & 0.999993919 &0.134372 $\frac{\rm arcs}{\rm revolution}$ 
& $9.05\times 10^8$ & 93.9\\
S2 & 5693.30424 & 0.999979485 & 6.01576 $\frac{\rm arcs}{\rm revolution}$
&$3.26\times 10^6$ &15.15 \\
S8 & 9238.88965 & 0.999992385 & 2.90826 $\frac{\rm arcs}{\rm revolution}$ &$2.98\times 10^7$ & 67.0\\
S12 & 10641.9649 & 0.999991213 & 2.35213  $\frac{\rm arcs}{\rm revolution}$ &$2.978\times 10^7$ & 54.05\\
S13 & 36897.1620 & 0.999988560 & 0.364077 $\frac{\rm arcs}{\rm revolution}$
& $1.295\times 10^8$ & 36.4 \\
S14 & 5321.06355 & 0.999988863 & 6.65917 $\frac{\rm arcs}{\rm revolution}$&$7.37\times 10^6$ & 37.9 \\
\hline \hline
\end{tabular}
\end{center}
\caption{Predictions for Lense-Thirring effect $\Delta\phi^{\rm {GTR}}$, 
LTP and orbital period $\rm P$  for the observed orbits of 
  S-stars around the rotating galactic centre black hole. The Kerr parameter 
was chosen to be $a_{\rm Galactic}=0.9953\frac{GM_{\rm BH}}{c^2}$. The values 
of Carter's constant $Q$ are in units of $\frac{G^2 M_{\rm BH}^2}{c^4}$.}
\label{EINNPolarNStime2hs1}
\end{table}

\begin{table}
\begin{center}
\begin{tabular}{|c|c|c|c|c|c||}\hline\hline
$\bf {star}$ &  $Q$ & $E$ & $\Delta\phi^{\rm GTR}$ & LTP(yr) & ${\rm P}$(yr)\\ \hline
S1 & 71705.3106 & 0.999993919 &0.134978 $\frac{\rm arcs}{\rm revolution}$ 
& $9.01\times 10^8$ & 93.9\\
S2 & 5693.30424 & 0.999979485 & 6.042901 $\frac{\rm arcs}{\rm revolution}$
&$3.25\times 10^6$ &15.15 \\
S8 & 9238.88965 & 0.999992385 & 2.92138 $\frac{\rm arcs}{\rm revolution}$ &$2.97\times 10^7$ & 67.0\\
S12 & 10641.9649 & 0.999991213 & 2.36275  $\frac{\rm arcs}{\rm revolution}$ &$2.96\times 10^7$ & 54.05\\
S13 & 36897.1620 & 0.999988560 & 0.36572 $\frac{\rm arcs}{\rm revolution}$
& $1.29\times 10^8$ & 36.4 \\
S14 & 5321.06355 & 0.999988863 & 6.68921 $\frac{\rm arcs}{\rm revolution}$&$7.34\times 10^6$ & 37.9 \\
\hline \hline
\end{tabular}
\end{center}
\caption{Predictions for Lense-Thirring effect $\Delta\phi^{\rm {GTR}}$, 
LTP and orbital period $\rm P$  for the observed orbits of 
  S-stars around the rotating galactic centre black hole. The Kerr parameter 
was chosen to be $a_{\rm Galactic}=0.99979\frac{GM_{\rm BH}}{c^2}$. The values 
of Carter's constant $Q$ are in units of $\frac{G^2 M_{\rm BH}^2}{c^4}$.}
\label{EINNPolarNStime2hs2}
\end{table}

\begin{table}
\begin{center}
\begin{tabular}{|c|c|c|c|c|c||}\hline\hline
$\bf {star}$ &  $Q$ & $E$ & $\Delta\phi^{\rm GTR}$ & LTP(yr) & ${\rm P}$(yr)\\ \hline
S1 & 65431.43821  & 0.999993545  & 0.15394$\frac{\rm arcs}{\rm revolution}$ 
& $7.22\times 10^8$ & 85.8 \\
S2 & 5273.53220& 0.999979145 & 6.73964 $\frac{\rm arcs}{\rm revolution}$
&$2.84\times 10^6$ &14.8 \\
S8 & 6524.03518 & 0.999991945 & 4.89646 $\frac{\rm arcs}{\rm revolution}$ &$1.63\times 10^7$ & 61.6\\
S12 & 9730.48589  &0.999990853 & 2.68671  $\frac{\rm arcs}{\rm revolution}$ & 
$2.45\times 10^7$ & 50.9\\
S13 & 26279.90639  & 0.99998444& 0.604888$\frac{\rm arcs}{\rm revolution}$
&$4.91\times 10^7$ & 22.9 \\
S14 & 4204.76359 & 0.999987653 & 9.47151 $\frac{\rm arcs}{\rm revolution}$&$4.43\times 10^6$ & 32.4 \\
\hline \hline
\end{tabular}
\end{center}
\caption{Predictions for Lense-Thirring effect $\Delta\phi^{\rm {GTR}}$, 
LTP and orbital period $\rm P$ for the observed orbits of 
  S-stars around the rotating galactic centre black hole. The Kerr parameter 
was chosen to be $a_{\rm Galactic}=0.9939\frac{GM_{\rm BH}}{c^2}$. The values 
of Carter's constant $Q$ are in units of $\frac{G^2 M_{\rm BH}^2}{c^4}$.}
\label{EINNPolarNStime3a}
\end{table}

We now proceed to calculate, using our exact analytic 
solutions,  the Lense-Thirring effect and the corresponding Lense-Thirring period for the observed orbits of S-stars: S1,S2,S8,S12,
S13,S14 for 
various values of the Kerr parameter. These orbits have 
been measured by Eisenhauer et al \cite{SINFONI} at near-infrared, with 
the data taken with the new adaptive optics-assisted integral-field 
spectrometer SINFONI on the European Southern Observatory (ESO) 
Very-Large-Telescope (VLT).
Orbital data for the semi-major axis ${\bf a}$  and the eccentricity 
$
{\bf e}$ of the corresponding 
Keplerian 
orbits are listed in \cite{SINFONI}.
Our motivation for this application was to investigate if the dragging of the orbital plane by the Lense-Thirring effect can help in understanding the observed distribution of the 
orbital planes for the S-stars. 
The parameters of the exact theory of non-spherical polar orbits 
developed in the previous sections are the invariant parameters $Q,E$, 
the spin and the mass of the 
galactic black hole, $a$, $M_{\rm BH}$ respectively.
For the black hole mass we adopt 
a value: $M_{\rm BH}=4.06\times 10^6 M_{\odot}$.
The choice of values for the invariant parameters is restricted by 
requiring that 
the predictions of the theory for the periastron, apoastron distances and 
the period $\rm P$ are 
in agreement with the orbital data in \cite{SINFONI}. 
We present our results for four 
different values of the Kerr parameter in Tables \ref{EINNPolarNStime1}-\ref{EINNPolarNStime2hs2} \footnote{In tables \ref{EINNPolarNStime2hs1},\ref{EINNPolarNStime2hs2} we have 
calculated the Lense-Thirring precessions for the values of the spin 
of the black hole for which the Aschenbach effect (hump of 
the LNRF-velocity profile) becomes to be relevant 
a=0.9953 for Keplerian discs \cite{Aschenbach} and a=0.99979 for  marginallly stable discs respectively \cite{ZdenekA}.} 
 respectively \footnote{The 
periapsis advance for the S-stars with the initial conditions as those 
in tables \ref{EINNPolarNStime1}, \ref{EINNPolarNStime2} have been calculated, using the exact closed 
formula eq.(\ref{PERIASTRONSHIFTP}), in appendix \ref{PeriapsisBH}, table \ref{PeriapsisPclosedform}.} .
Indeed as we can see the stars S2 and S14 have LTP's of the right size as compared with estimations of their age from 
spectroscopic analysis and the average stellar ages in the two outer
 star rings/discs \cite{Paumard}. According to \cite{Paumard} it is 
estimated that the stellar discs in the galactic centre have formed about 
$6\pm 2$ Myr ago.
Thus S2 and S14 because their LTPs are comparable to the 
average ages of the stellar contents of the discs 
could have originated from the stellar discs 
before the Lense-Thirring dragging brought them where they are currently 
observed. The LTP's decrease with higher values of the 
spin 
of the black hole. Our results in Table \ref{EINNPolarNStime1} agree 
with the results in \cite{SINFONI}, where the Lense-Thirring periods were 
calculated for $a=0.52$ using a weak field approximation formula 
\cite{LB}. This is 
a nice check of the weak field limit of the exact theory.

We repeated the analysis for $a=0.9939$ but for different values of the 
invariant parameters $Q,E$. This choice is still consistent with experimental 
data for the semi-major axis ${\bf a}$ and eccentricity ${\bf e}$ quoted 
in \cite{SINFONI}. The results are presented in Table \ref{EINNPolarNStime3a}.
Now we see that also the LTP for S8 is of the right magnitude. Thus we 
can conclude that the Lense-Thirring effect can help in understanding the 
observed distribution of the orbital planes for some of the S-stars but 
perhaps for not all of them \footnote{See \cite{Paumard} 
for a recent summary of attempts to explain the youth of S-stars.}. On the other hand when more precise 
astrometric data will become  available for all of the S-stars one should 
 try to fit the 
experimental raw data using the relativistic geodesic equations of motion and the exact results obtained in this work, 
assuming that 
the galactic centre is a Kerr black hole, rather than try to fit the data 
with Keplerian orbits.

Our choice of value for the black hole mass is consistent 
with the best fit central mass (for a distance $R_0$ to the 
galactic centre of 8Kpc) in \cite{SINFONI}. If we try to fit the data 
for the periastron and apoastron distances in \cite{SINFONI} with a smaller 
black hole mass we find that the orbital period P increases. In particular, 
adopting the value $M_{\rm BH}=3.59\times 10^6 M_{\odot}$  reproducing 
the measured values for $r_P,r_A$ \cite{SINFONI} for the star S2 results in P$>15.6$yr 
in conflict with experiment \cite{SINFONI}. 

\section{Zooming in on the galactic centre black hole}
\subsubsection{Frame dragging and periapsis advance of 
non-spherical polar orbits in the central milliarcsecond of Sgr A$^{*}$}
\label{MILLITOKSODEUTEROLEPTOU}
In this subsection we shall apply our exact formulae for the calculation 
of the relativistic periastron advance and frame dragging for non-spherical 
polar orbits in the central milliarcsecond of the galactic centre of Milky Way.
These cusp orbits are the target of the GRAVITY experiment and will provide 
an interesting enviroment for probing stellar dynamics in the strong-field 
regime of General Relativity \footnote{GRAVITY is an adaptive optics (AO) 
assisted, near-infrared VLTI instrument for precision narrow-angle astrometry 
and interferometric phase referenced imaging of faint objects. With an 
accuracy of 10 micro-arcseconds (10$\mu$as), GRAVITY will be able to study 
motions to within a few times the event horizon size of the Galactic centre 
black hole.}.

Assuming that the centre of Milky Way is a rotating black hole and the 
spacetime near the region Sgr A$^{*}$ is described by the Kerr 
geometry, we calculated using  (\ref{TimelikePolar}) the precise 
frame dragging of 
a timelike stellar orbit with a non-spherical polar geometry. 
The results are displayed 
in tables \ref{EINSTEINPolarNStime} and \ref{EINSTEINPolarNStime1}.
The periods for the orbits in table \ref{EINSTEINPolarNStime1} were 
calculated to be: $6.443\times 10^{3}$ s, $6.359\times 10^3$ s for 
$a=0.52\frac{GM_{\rm {BH}}}{c^2},0.9939\frac{GM_{\rm {BH}}}{c^2}$ respectively, assuming a galactic black hole mass 
$M_{\rm BH}=4.060\times 10^6 M_{\odot}$.
\begin{table}
\begin{center}
\begin{tabular}{|c|c|c|c|}\hline\hline
{\bf parameters} & {\bf Periastron} & {\bf Apoastron} & {\bf predicted dragging} \\
$a_{\rm Galactic}=0.9939,$ & 5.51905 & 14.5373 & $\Delta\phi^{\rm GTR}=1.08552=223905 \frac{\rm arcs}{\rm revolution}$ \\
 $E=0.956853,Q=13$ & & & =$62.19^{\circ} {\rm per\; revolution}$ \\
$a_{\rm Galactic}=0.52, $ & 4.80864 & 14.5620 &  $\Delta\phi^{\rm GTR}=1.0349=213464 \frac{\rm arcs}{\rm revolution}$ \\
$E=0.956853,Q=13$& & & =$59.29^{\circ}{\rm per\; revolution}$ \\
\hline \hline
\end{tabular}
\end{center}
\caption{Predictions for frame dragging $\Delta\phi^{\rm {GTR}}$ from galactic black hole for a timelike 
non-spherical polar stellar orbit for two different values of the Kerr 
parameter $a_{\rm Galactic}$. Carter's constant is fixed at  $Q=13 \frac{G^2 M_{{\rm BH}}^2}{c^4}$ while $E=0.956853$.
 The values of the radii and Kerr parameter are 
in units of $GM_{{\rm {BH}}}/c^2$.}
\label{EINSTEINPolarNStime}
\end{table}

\begin{table}
\begin{center}
\begin{tabular}{|c|c|c|c|}\hline\hline
{\bf parameters} & {\bf Periastron} & {\bf Apoastron} & {\bf predicted dragging} \\
$a_{\rm Galactic}=0.9939,$ & 7.90470 & 12.5692 & $\Delta\phi^{\rm GTR}=0.6477498=133608 \frac{\rm arcs}{\rm revolution}$ \\
 $E=0.956853,Q=14$ & & & =$37.11^{\circ} {\rm per\; revolution}$ \\
$a_{\rm Galactic}=0.52, $ & 7.66620 & 12.6339 &  $\Delta\phi^{\rm GTR}=0.366361=75567.5 \frac{\rm arcs}{\rm revolution}$ \\
$E=0.956853,Q=14$& & & =$20.991^{\circ}{\rm per\; revolution}$ \\
\hline \hline
\end{tabular}
\end{center}
\caption{Predictions for frame dragging from galactic black hole for timelike 
non-spherical polar stellar orbit for the same values for the Kerr parameter 
and $E$ as in table \ref{EINSTEINPolarNStime} but for a value of 
Carter's constant $Q=14\frac{G^2 M_{{\rm BH}}^2}{c^4}$.
 The values of the radii and Kerr parameter are 
in units of $GM_{{\rm BH}}/c^2$.}
\label{EINSTEINPolarNStime1}
\end{table}

Let us give subsequently  an example of a non-spherical stellar polar orbit of low eccentricity. The results are exhibited in table 
\ref{EINSTEINPolarNStime2}.

\begin{table}
\begin{center}
\begin{tabular}{|c|c|c|c|}\hline\hline
{\bf parameters} & {\bf Periastron} & {\bf Apoastron} & {\bf predicted dragging} \\
$a_{\rm Galactic}=0.99979,$ & 107.703 & 112.222 & $\Delta\phi^{\rm GTR}=0.0112114=2312.52 \frac{\rm arcs}{\rm revolution}$ \\
$E=0.9954853,Q=113$ & & & =$0.642368^{\circ} {\rm per\; revolution}$ \\
$a_{\rm Galactic}=0.99616,$ & 107.703 & 112.222 & $\Delta\phi^{\rm GTR}=0.0111707=2304.13 \frac{\rm arcs}{\rm revolution}$ \\
$E=0.9954853,Q=113$ & & & =$0.640036^{\circ} {\rm per\; revolution}$ \\
$a_{\rm Galactic}=0.9953,$ & 107.703 & 112.222 & $\Delta\phi^{\rm GTR}=0.0111611=2302.14 \frac{\rm arcs}{\rm revolution}$ \\
$E=0.9954853,Q=113$ & & & =$0.639484^{\circ} {\rm per\; revolution}$ \\
$a_{\rm Galactic}=0.9939,$ & 107.703 & 112.222 & $\Delta\phi^{\rm GTR}=0.0111454=2298.9 \frac{\rm arcs}{\rm revolution}$ \\
 $E=0.9954853,Q=113$ & & & =$0.638584^{\circ} {\rm per\; revolution}$ \\
$a_{\rm Galactic}=0.52, $ & 107.697 & 112.227 &  $\Delta\phi^{\rm GTR}=0.00583177=1202.89 \frac{\rm arcs}{\rm revolution}$ \\
$E=0.9954853,Q=113$& & & =$0.334136^{\circ}{\rm per\; revolution}$ \\
\hline \hline
\end{tabular}
\end{center}
\caption{Predictions for frame dragging from a galactic black hole for timelike 
non-spherical polar stellar {\em low-eccentricity} orbits.
 The values of the radii and Kerr parameter are 
in units of $GM_{{\rm BH}}/c^2$, 
while those of Carter's constant $Q$ in 
units of $(GM_{{\rm BH}}/c^2)^2$.}
\label{EINSTEINPolarNStime2}
\end{table}

Using the exact formula, equation (\ref{PERIASTRONSHIFTP}), the periastron advance for the orbits with initial conditions displayed in 
table \ref{EINSTEINPolarNStime2} were calculated.  Our results 
are summarized in table \ref{EINSTEINPolarNStimePAD}.
Thus, the closer we get to the galactic centre the relativistic effect 
of periapsis advance becomes very substantial. We observe that the 
effect increases with decreasing Kerr parameter.
The semi-major axis $\alpha=\frac{r_P+r_A}{2}\sim 110 \frac{GM_{\rm BH}}{c^2}=
55 R_s$ corresponds to an angle of observation $\sim 0.54 mas=540 \mu as$ 
(1mas$\equiv 10^{-3}$ arcs)\footnote{We assume a black hole mass 
$M_{\rm BH}=4.06\times 10^6 M_{\odot}$ and a galactic centre distance of 
$8.0$Kpc.}. The calculated orbital period ${\rm P}$ is 1.72 days. Measurement of 
such orbits and the corresponding relativistic effects 
close to the outer event horizon $r_+$ 
of the candidate galactic black hole can 
provide an important test of the theory of general relativity at the strong 
field regime.
\begin{table}
\begin{center}
\begin{tabular}{|c|c|c|c|}\hline\hline
{\bf parameters} & {\bf $r_P$} & {\bf $r_A$} & {\bf predicted periastron advance} \\
$a_{\rm Galactic}=0.99979,$ & 107.703 & 112.222 & $\Delta\Psi^{\rm GTR}=
0.178388=36795.2 \frac{\rm arcs}{\rm revolution}$ \\
$E=0.9954853,Q=113$ & & & =$10.2209^{\circ} {\rm per\; revolution}$ \\

$a_{\rm Galactic}=0.99616,$ & 107.703 & 112.222 & $\Delta\Psi^{\rm GTR}=
0.178391=36795.9 \frac{\rm arcs}{\rm revolution}$ \\
$E=0.9954853,Q=113$ & & & =$10.2211^{\circ} {\rm per\; revolution}$ \\
$a_{\rm Galactic}=0.9953,$ & 107.703 & 112.222 & $\Delta\Psi^{\rm GTR}=
0.178392=36796.0 \frac{\rm arcs}{\rm revolution}$ \\
$E=0.9954853,Q=113$ & & & =$10.2211^{\circ} {\rm per\; revolution}$ \\
$a_{\rm Galactic}=0.9939,$ & 107.703 & 112.222 & $\Delta\Psi^{\rm GTR}=0.178393=36796.3 \frac{\rm arcs}{\rm revolution}$ \\
 $E=0.9954853,Q=113$ & & & =$10.2212^{\circ} {\rm per\; revolution}$ \\
$a_{\rm Galactic}=0.52, $ & 107.697 & 112.227 &  $\Delta\Psi^{\rm GTR}=0.178724=36864.4 \frac{\rm arcs}{\rm revolution}$ \\
$E=0.9954853,Q=113$& & & =$10.2401^{\circ}{\rm per\; revolution}$ \\
\hline \hline
\end{tabular}
\end{center}
\caption{Predictions for periastron advance from a galactic black hole for timelike 
non-spherical, polar, stellar low eccentricity orbit.
 The values of the radii and Kerr parameter are 
in units of $GM_{{\rm BH}}/c^2$, 
while those of Carter's constant $Q$ in 
units of $(GM_{{\rm BH}}/c^2)^2$.}
\label{EINSTEINPolarNStimePAD}
\end{table}

\begin{table}
\begin{center}
\begin{tabular}{|l|c|c|}\hline\hline
{\bf Orbital parameters} & {\bf Case 1: a=0.52} & {\bf Case 2: a=0.99616}  \\
$r_P$ & 207.628 $\frac{GM_{\rm BH}}{c^2}$  & 207.628 $\frac{GM_{\rm BH}}{c^2}$\\
$r_A$ & 2005.837 $\frac{GM_{\rm BH}}{c^2}$  & 2005.837 $\frac{GM_{\rm BH}}{c^2}$\\
Periapsis advance(arcs/rev.) &10460.03 & 10454.7 \\
Frame dragging (arcs/rev) & 186.626 & 357.513 \\
Periapsis advance ($\frac{\circ}{\rm yr}$) \footnotemark[13]{
Assumes $M_{\rm BH}=4.06\times 10^6 M_{\odot}$}
  & 19.8  & 19.8 \\
L.T. effect ($\frac{\circ}{\rm yr}$) & 0.353 & 0.675\\
Orbital Period $\rm P$ (days) & 53.6 & 53.6\\
L.T. Period (yr) & $1.02 \times 10^3$ & 533\\
\hline \hline
\end{tabular}
\end{center}
\caption{Calculation of orbital data and 
the corresponding relativistic effects for a cusp polar 
non-spherical stellar orbit, around the galactic 
black hole, for two different values of the 
Kerr parameter with $Q=380 \frac{G^2 M^2_{\rm BH}}{c^4},E=0.99954853$.
 }
\label{CuspOrbit}
\end{table}

A very interesting example of a stellar polar non-spherical orbit is 
presented in Table \ref{CuspOrbit}. We present the orbital parameters 
for two choices of the spin of the black hole. The rest of the initial 
conditions are $Q=380\frac{G^2 M^2_{\rm BH}}{c^4}, E=0.99954853$.
The angular resolution of the periapsis $r_P$ is $\sim 1$mas. 
We see from the results in the table that the relativistic phenomenon of 
periapsis precession is quite substantial and it amounts to 
$\frac{19.8^{\circ}}{\rm yr}$ with the orbital period of only 53.6 days. 
For the convenience of the reader we also list the periapsis precession 
and Lense-Thirring effect after 
one revolution as well as the Lense-Thirring period. As we mentioned, 
VLBI observations aim to detect such cusp stellar orbits which are closer 
to the candidate black hole than the observed S-stars. 
The periapsis precession in the example we discuss should be measured with 
a good precision after one year of observations. We repeated the analysis 
for higher values of the spin of the black hole and the results are 
dislayed in table \ref{CuspOrbitAS}.

\begin{table}
\begin{center}
\begin{tabular}{|l|c|c|}\hline\hline
{\bf Orbital parameters} & {\bf a=0.9953} & {\bf  a=0.99979}  \\
$r_P$ & 207.628 $\frac{GM_{\rm BH}}{c^2}$  & 207.628 $\frac{GM_{\rm BH}}{c^2}$\\
$r_A$ & 2005.837 $\frac{GM_{\rm BH}}{c^2}$  & 2005.837 $\frac{GM_{\rm BH}}{c^2}$\\
Periapsis advance(arcs/rev.) &10454.8 & 10454.7 \\
Frame dragging (arcs/rev) & 357.204 & 358.815 \\
Periapsis advance ($\frac{\circ}{\rm yr}$) \footnotemark[13]{
Assumes $M_{\rm BH}=4.06\times 10^6 M_{\odot}$}
  & 19.8  & 19.8 \\
L.T. effect ($\frac{\circ}{\rm yr}$) & 0.675 & 0.679\\
Orbital Period $\rm P$ (days) & 53.6 & 53.6\\
L.T. Period (yr) & $533$ & 531\\
\hline \hline
\end{tabular}
\end{center}
\caption{Calculation of orbital data and 
the corresponding relativistic effects for a cusp polar 
non-spherical stellar orbit, around the galactic 
black hole, for two different values of the 
Kerr parameter with $Q=380 \frac{G^2 M^2_{\rm BH}}{c^4},E=0.99954853$.
 }
\label{CuspOrbitAS}
\end{table}

One might also wonder about the tidal effects on the orbits of the 
stars we just discussed. We can estimate these effects as follows: 
tidal disruption of the star with mass 
$m_{*}$ and radius $R_{*}$ is expected to occur only if the star approaches 
the black hole to within its Roche radius
\begin{equation}
r_{\rm R}=\left(\frac{M_{\rm BH}}{m_{*}}\right)^{1/3} R_{*}
\end{equation}
For a star like our Sun with $R_{*}=R_{\odot}$ and $m_{*}=M_{\odot}$ 
and for a galactic black hole mass $M_{\rm BH}=4.06\times 10^6
M_{\odot}$ the Roche radius is calculated to be $r_{\rm R}=18.5 
\frac{GM_{\rm BH}}{c^2}$. Thus the ratio of the periastron radius to the 
Roche radius for the orbits presented in tables \ref{EINSTEINPolarNStimePAD},\ref{CuspOrbitAS} is 5.8 and 11.2 respectively. Consequently, any tidal 
effects are small for a solar-type star in orbit around the galactic centre 
black hole according to the data in tables  \ref{EINSTEINPolarNStimePAD}, \ref{CuspOrbitAS} and we do not expect significant mass loss. 
If a star similar to S2 with $m_{*}=15M_{\odot}$,  $R_{*}=7 R_{\odot}$ 
is in a polar orbit in the central mas of Sgr A$^*$ according to the data in tables  \ref{EINSTEINPolarNStimePAD}, \ref{CuspOrbitAS} the ratio $\frac{r_P}{r_{\rm R}}$ is calculated to be 
2.05 and 3.95 respectively.
For a neutron star with mass $\sim 1.4 M_{\odot}$ and radius 
$\sim 10$Km in a non-spherical polar trajectory  with the orbital 
characteristics of tables \ref{EINSTEINPolarNStimePAD}-\ref{CuspOrbitAS} 
the tidal effects are negligible. 

On the other hand in order to investigate 
more precisely  the tidal interaction of a star in a 
high eccentricity orbit around the supermassive galactic centre black hole 
a full relativistic treatment is required. Such an analysis which will involve 
the calculation of the tidal tensor for non-spherical polar, equatorial or
arbitrary orbital inclination test particle trajectories with or without 
the cosmological constant is however beyond the scope of the current work.
It is a promising avenue for future research since the tidal interaction 
of a star by supermassive black hole may lead to observable effects \cite{JMR}.

\section{Periastron precession of non-circular equatorial orbits 
around a central rotating mass in the presence of  a cosmological 
constant}
\label{PeriastronLambdaAdvance}

We now proceed to derive exact expressions for the periastron advance and 
orbital period for a timelike non-circular orbit in the Kerr field  
in the presence of a cosmological constant.
In the former case the  relevant differential equation is \footnote{Setting $Q=0, \theta=\pi/2$ in the relevant equations in equations \ref{LARTH}.}
\begin{equation}
\frac{d\phi}{dr}=\frac{-\Xi^2(a E-L)}{\sqrt{R}}+
\frac{a\Xi^2}{\Delta_r \sqrt{R}}\left[(r^2+a^2)E-aL\right]
\label{cosmologicalbh}
\end{equation}
In (\ref{cosmologicalbh}) $R$ is defined in (\ref{LARTH}) with $Q=0$.

For the derivation of an analytical exact expression of periastron advance 
for the 
equatorial orbit of a test particle in Kerr-de Sitter spacetime we need 
to calculate the following Abelian integrals
\begin{equation}
\Delta\phi=2\left[\int_{r_A}^{r_P}\frac{-\Xi^2(a E-L)}{\sqrt{R}}dr+
\int_{r_A}^{r_P}\frac{a\Xi^2}{\Delta_r \sqrt{R}}\left[(r^2+a^2)E-aL\right]dr\right]
\label{TotalLambdaIntegral}
\end{equation}
Let us now calculate the above Abelian integrals.
The first integral is in the general case (when all roots are 
distinct) a hyperelliptic integral. 
By applying the transformation 
\begin{equation}
z=\frac{\alpha_{\mu-1}-\alpha_{\mu+1}}{\alpha_{\mu}-\alpha_{\mu+1}}
\frac{r-\alpha_{\mu}}{r-\alpha_{\mu-1}}
\end{equation}
and organizing the roots of the polynomial $R$ 
as follows 
\begin{equation}
\alpha_{\nu}>\alpha_{\mu}>\alpha_{\rho}>\alpha_i
\end{equation}
where $\alpha_{\nu}=\alpha_{\mu-1}, \alpha_{\rho}=\alpha_{\mu+1}=r_P, 
\alpha_{\mu}=r_A, \alpha_i=\alpha_{\mu+i+1},i=1,2,3$, 
our genus 2 hyperelliptic integral is reduced to the 
integral representation of Lauricella's fourth, generalized hypergeometric 
function of three variables $F_D$ \cite{LAURICELLA} \footnote{See 
Appendix \ref{FDLauricella} for its definition, the differential equations 
it obeys and its integral representation.}.
More precisely we obtain

\begin{eqnarray}
\int_{r_A}^{r_P}\frac{-\Xi^2(a E-L)}{\sqrt{R}}dr&=&
\frac{-\Xi^2 (a E-L)}{\sqrt{\frac{\Lambda}{3}}H}\sqrt{\omega}\Biggl[
-F_D\left(\frac{1}{2},\frac{1}{2},\frac{1}{2},\frac{1}{2},1,\kappa^2,
\lambda^2,\mu^2 \right)\frac{\Gamma(1/2)^2}{\Gamma(1)} \nonumber \\
&+&\omega F_D\left(\frac{3}{2},\frac{1}{2},\frac{1}{2},\frac{1}{2},2,\kappa^2,\lambda^2,\mu^2\right)\frac{\Gamma(\frac{3}{2})\Gamma(\frac{1}{2})}{\Gamma(2)}
\Biggr] \nonumber \\
\label{HyperLauricella}
\end{eqnarray} 
where 
\begin{equation}
H:=\sqrt{(\alpha_{\mu}-\alpha_{\mu+1})(\alpha_{\mu}-\alpha_{\mu+2})
(\alpha_{\mu}-\alpha_{\mu+3})(\alpha_{\mu}-\alpha_{\mu+4})}
\label{CAPH}
\end{equation}
and $\omega$ is now defined as
\begin{equation}
\omega:=\frac{\alpha_{\mu}-\alpha_{\mu+1}}{\alpha_{\mu-1}-\alpha_{\mu+1}}
\label{OMEGAL}
\end{equation}
In addition we have defined the dimensionless quantity 
$\Lambda$ from the relation 
$\Lambda:=${\boldmath$\Lambda$}$(\frac{GM_{\rm BH}}{c^2})^2$. 
The variables of the hypergeometric function $F_D$ are given by the 
expressions
\begin{eqnarray}
\kappa^2 &=& \frac{\alpha_{\mu}-\alpha_{\mu+1}}{\alpha_{\mu-1}-\alpha_{\mu+1}}
\frac{\alpha_{\mu-1}-\alpha_{\mu+2}}{\alpha_{\mu}-\alpha_{\mu+2}}=
\omega\frac{\alpha_{\mu-1}-\alpha_{\mu+2}}{\alpha_{\mu}-\alpha_{\mu+2}}=
\frac{\alpha-\beta}{r_{\Lambda}^1-\beta}\frac{r_{\Lambda}^1-\gamma}{\alpha-\gamma} \nonumber \\
\lambda^2&=&\omega\frac{\alpha_{\mu-1}-\alpha_{\mu+3}}{\alpha_{\mu}-\alpha_{\mu+3}}=\frac{\alpha-\beta}{r_{\Lambda}^1-\beta}\frac{r_{\Lambda}^1}{\alpha} \nonumber \\
\mu^2 &=& \omega\frac{\alpha_{\mu-1}-\alpha_{\mu+4}}{\alpha_{\mu}-\alpha_{\mu+4}}=\frac{\alpha-\beta}{r_{\Lambda}^1-\beta}\frac{r_{\Lambda}^1-r_{\Lambda}^2}{\alpha-r_{\Lambda}^2} \nonumber \\
\label{ARGULAURI}
\end{eqnarray}

Now for the second integral in (\ref{TotalLambdaIntegral}) 
we use partial fraction expansion
\begin{equation}
\frac{a\Xi^2 \left[(r^2+a^2)E-aL\right]}{\Delta_r}=
\frac{A^1}{r-r_{\Lambda}^+}+\frac{A^2}{r-r_{\Lambda}^-}+
\frac{A^3}{r-r_+}+\frac{A^4}{r-r_-}
\end{equation}
For the coefficients $A^i, i=1,\cdots 4$ we obtain the expressions
\begin{eqnarray}
A^1&=&\frac{-3 a^2 \Xi^2 \left[a E -L+E\alpha_1^2\right]}{
(\alpha_1-\beta_1)(\alpha_1-\gamma_1)(\alpha_1-\delta_1)\Lambda} \nonumber \\
A^2&=&\frac{3 a^2 \Xi^2 \left[a E -L+E\delta_1^2\right]}{
(\alpha_1-\delta_1)(\delta_1-\beta_1)(\delta_1-\gamma_1)\Lambda} \nonumber \\
A^3&=&\frac{3 a^2 \Xi^2 \left[a E -L+E\beta_1^2\right]}{
(-\alpha_1+\beta_1)(-\beta_1+\gamma_1)(\beta_1-\delta_1)\Lambda} \nonumber \\
A^4&=&\frac{-3 a^2 \Xi^2 \left[a E -L+E\gamma_1^2\right]}{
(\alpha_1-\gamma_1)(\beta_1-\gamma_1)(\gamma_1-\delta_1)\Lambda} \nonumber \\
\end{eqnarray}
where we have defined
\begin{equation}
r_{\Lambda}^+=\alpha_1,r_+=\beta_1,r_-=\gamma_1,r_{\Lambda}^-=\delta_1
\end{equation}
for the roots of the polynomial $\Delta_r$.
Proceeding in a similar way as before we transform the Abelian integrals 
into the familiar integral representation of Lauricella's function $F_D$.
For instance for the integral $\int_{r_A}^{r_P}\frac{A_4}{(r-r_-)}\frac{dr}{\sqrt{R}}$ we derive an exact expression in terms of Lauricella's function $F_D$ of four variables:
\begin{eqnarray}
\int_{r_A}^{r_P}\frac{A_4}{(r-r_-)}\frac{dr}{\sqrt{R}}&=& \frac{A_4\sqrt{\omega}}{\sqrt{\frac{\Lambda}{3}}H_{-}}\Biggl[
F_D\left(\frac{1}{2},\frac{1}{2},1,\frac{1}{2},\frac{1}{2},1,
\kappa^2,\lambda^2,\mu^2,\nu^2\right)\frac{\Gamma^2\left(\frac{1}{2}\right)}{
\Gamma(1)}\nonumber \\
&+&\omega^2 F_D \left(\frac{5}{2},\frac{1}{2},1,\frac{1}{2},\frac{1}{2},3,\kappa^2,\lambda^2,\mu^2,\nu^2\right)\frac{\Gamma\left(\frac{5}{2}\right)\Gamma\left(\frac{1}{2}\right)}{\Gamma(3)}\nonumber \\
&-&2 \omega F_D\left(\frac{3}{2},\frac{1}{2},1,\frac{1}{2},\frac{1}{2},2,
\kappa^2,\lambda^2,\mu^2,\nu^2\right)\frac{\Gamma\left(\frac{3}{2}\right)\Gamma\left(\frac{1}{2}\right)}{\Gamma(2)}\Biggr] \nonumber \\
\label{FourRminus}
\end{eqnarray}
In equation (\ref{FourRminus}) the four moduli are given in terms of the 
roots of the polynomial $R$ and the radii of the black hole horizons 
by the expressions
\begin{eqnarray}
\kappa^2&=&\frac{\alpha_{\mu}-\alpha_{\mu+1}}{\alpha_{\mu-1}-\alpha_{\mu+1}}
\frac{\alpha_{\mu+2}-\alpha_{\mu-1}}{\alpha_{\mu+2}-\alpha_{\mu}}
=\frac{\alpha-\beta}{r_{\Lambda}^1-\beta}\frac{\gamma-r_{\Lambda}^1}{\gamma-
\alpha} \nonumber \\
\lambda^2&=&\frac{\alpha_{\mu}-\alpha_{\mu+1}}{\alpha_{\mu-1}-\alpha_{\mu+1}}
\frac{\alpha_{\mu+3}-\alpha_{\mu-1}}{\alpha_{\mu+3}-\alpha_{\mu}}=
\frac{\alpha-\beta}{r_{\Lambda}^1-\beta}\frac{r_{-}-r_{\Lambda}^1}{r_{-}-
\alpha} \nonumber \\
\mu^2&=&\frac{\alpha_{\mu}-\alpha_{\mu+1}}{\alpha_{\mu-1}-\alpha_{\mu+1}}
\frac{\alpha_{\mu-1}}{\alpha_{\mu}}=\frac{\alpha-\beta}{r_{\Lambda}^1-\beta}\frac{r_{\Lambda}^1}{\alpha} \nonumber \\
\nu^2&=&\frac{\alpha_{\mu}-\alpha_{\mu+1}}{\alpha_{\mu-1}-\alpha_{\mu+1}}
\frac{\alpha_{\mu+6}-\alpha_{\mu-1}}{\alpha_{\mu+6}-\alpha_{\mu}}=
\frac{\alpha-\beta}{r_{\Lambda}^1-\beta}\frac{r_{\Lambda}^2-r_{\Lambda}^1}{
r_{\Lambda}^2-\alpha} \nonumber \\
\end{eqnarray}
while $H_{-}$ is defined to be:
\begin{eqnarray}
H_{-}&:=&\sqrt{(\alpha_{\mu}-\alpha_{\mu+1})(\alpha_{\mu}-\alpha_{\mu+2})}
(\alpha_{\mu}-\alpha_{\mu+3})\sqrt{\alpha_{\mu}(\alpha_{\mu}-\alpha_{\mu+6})}
\nonumber \\
&=&\sqrt{(\alpha-\beta)(\alpha-\gamma)}(\alpha-r_{-})\sqrt{\alpha(
\alpha-r_{\Lambda}^2)}
\end{eqnarray}
In total we obtain:
\begin{eqnarray}
\Delta\phi^{\rm GTR}_{\rm E}&=&2\Biggl\{
\frac{-\Xi^2 (a E-L)}{\sqrt{\frac{\Lambda}{3}}H}\sqrt{\omega}
\nonumber \\ &\times& \Biggl[\nonumber 
-F_D\left(\frac{1}{2},\frac{1}{2},\frac{1}{2},\frac{1}{2},1,\frac{\alpha-\beta}{r_{\Lambda}^1-\beta}\frac{r_{\Lambda}^1-\gamma}{\alpha-\gamma},
\frac{\alpha-\beta}{r_{\Lambda}^1-\beta}\frac{r_{\Lambda}^1}{\alpha},
\frac{\alpha-\beta}{r_{\Lambda}^1-\beta}\frac{r_{\Lambda}^1-r_{\Lambda}^2}{\alpha-r_{\Lambda}^2} \right)\frac{\Gamma(1/2)^2}{\Gamma(1)} \nonumber \\
&+&\omega F_D\left(\frac{3}{2},\frac{1}{2},\frac{1}{2},\frac{1}{2},2,\frac{\alpha-\beta}{r_{\Lambda}^1-\beta}\frac{r_{\Lambda}^1-\gamma}{\alpha-\gamma},  
 \frac{\alpha-\beta}{r_{\Lambda}^1-\beta}\frac{r_{\Lambda}^1}{\alpha},
\frac{\alpha-\beta}{r_{\Lambda}^1-\beta}\frac{r_{\Lambda}^1-r_{\Lambda}^2}{\alpha-r_{\Lambda}^2}\right)\frac{\Gamma(\frac{3}{2})\Gamma(\frac{1}{2})}{\Gamma(2)}
\Biggr] \nonumber \\
&+&\frac{A^1\sqrt{\omega}}{\sqrt{\frac{\Lambda}{3}}H_{\Lambda}^+}
\nonumber \\
&\times & 
\Biggl[\omega^2 F_D\left(\frac{5}{2},1,\frac{1}{2},\frac{1}{2},\frac{1}{2},3,
\omega\frac{r_{\Lambda}^{+}-r_{\Lambda}^1}{r_{\Lambda}^{+}-\alpha},
\omega\frac{\gamma-r_{\Lambda}^1}{\gamma-\alpha},
\omega\frac{r_{\Lambda}^1}{\alpha},
\omega\frac{r_{\Lambda}^2-r_{\Lambda}^1}{r_{\Lambda}^2-\alpha}\right)\frac
{\Gamma\left(\frac{5}{2}\right)\Gamma\left(\frac{1}{2}\right)}{\Gamma(3)}\nonumber \\
&-&2 \omega F_D\left(\frac{3}{2},1,\frac{1}{2},\frac{1}{2},\frac{1}{2},2,
\omega\frac{r_{\Lambda}^{+}-r_{\Lambda}^1}{r_{\Lambda}^{+}-\alpha},
\omega\frac{\gamma-r_{\Lambda}^1}{\gamma-\alpha},
\omega\frac{r_{\Lambda}^1}{\alpha},
\omega\frac{r_{\Lambda}^2-r_{\Lambda}^1}{r_{\Lambda}^2-\alpha}\right)\frac
{\Gamma\left(\frac{3}{2}\right)\Gamma\left(\frac{1}{2}\right)}{\Gamma(2)} \nonumber \\
&+& F_D\left(\frac{1}{2},1,\frac{1}{2},\frac{1}{2},\frac{1}{2},1,
\omega\frac{r_{\Lambda}^{+}-r_{\Lambda}^1}{r_{\Lambda}^{+}-\alpha},
\omega\frac{\gamma-r_{\Lambda}^1}{\gamma-\alpha},
\omega\frac{r_{\Lambda}^1}{\alpha},
\omega\frac{r_{\Lambda}^2-r_{\Lambda}^1}{r_{\Lambda}^2-\alpha}\right)\frac
{\Gamma\left(\frac{1}{2}\right)\Gamma\left(\frac{1}{2}\right)}{\Gamma(1)}
\Biggr] \nonumber \\
&+&\frac{A^2\sqrt{\omega}}{\sqrt{\frac{\Lambda}{3}}H_{\Lambda}^-}
\nonumber \\
&\times & 
\Biggl[\omega^2 F_D\left(\frac{5}{2},\frac{1}{2},\frac{1}{2},\frac{1}{2},1,
3,\omega\frac{\gamma-r_{\Lambda}^1}{\gamma-\alpha},
\omega\frac{r_{\Lambda}^1}{\alpha},
\omega\frac{r_{\Lambda}^2-r_{\Lambda}^1}{r_{\Lambda}^2-\alpha},
\omega\frac{r_{\Lambda}^{-}-r_{\Lambda}^1}{r_{\Lambda}^{-}-\alpha}\right)
\frac{\Gamma\left(\frac{5}{2}\right)\Gamma\left(\frac{1}{2}\right)}
{\Gamma(3)}\nonumber \\
&-&2 \omega F_D \left(\frac{3}{2},\frac{1}{2},\frac{1}{2},\frac{1}{2},1,
2,\omega\frac{\gamma-r_{\Lambda}^1}{\gamma-\alpha},
\omega\frac{r_{\Lambda}^1}{\alpha},
\omega\frac{r_{\Lambda}^2-r_{\Lambda}^1}{r_{\Lambda}^2-\alpha},
\omega\frac{r_{\Lambda}^{-}-r_{\Lambda}^1}{r_{\Lambda}^{-}-\alpha}\right)
\frac{\Gamma\left(\frac{3}{2}\right)\Gamma\left(\frac{1}{2}\right)}
{\Gamma(2)}\nonumber \\
&+& F_D\left(\frac{1}{2},\frac{1}{2},\frac{1}{2},\frac{1}{2},1,
1,\omega\frac{\gamma-r_{\Lambda}^1}{\gamma-\alpha},
\omega\frac{r_{\Lambda}^1}{\alpha},
\omega\frac{r_{\Lambda}^2-r_{\Lambda}^1}{r_{\Lambda}^2-\alpha},
\omega\frac{r_{\Lambda}^{-}-r_{\Lambda}^1}{r_{\Lambda}^{-}-\alpha}\right)
\frac{\Gamma\left(\frac{1}{2}\right)\Gamma\left(\frac{1}{2}\right)}
{\Gamma(1)}\Biggr]
\nonumber \\
&+&\frac{A^3\sqrt{\omega}}{\sqrt{\frac{\Lambda}{3}}H_{+}}
\nonumber \\
&\times& \Biggl[\omega^2 F_D\left(\frac{5}{2},\frac{1}{2},1,\frac{1}{2},\frac{1}{2},3,
\omega\frac{\gamma-r_{\Lambda}^1}{\gamma-\alpha},
\omega\frac{r_{+}-r_{\Lambda}^1}{r_{+}-\alpha},
\omega\frac{r_{\Lambda}^1}{\alpha},
\omega\frac{r_{\Lambda}^2-r_{\Lambda}^1}{r_{\Lambda}^2-\alpha}\right)
\frac{\Gamma\left(\frac{5}{2}\right)\Gamma\left(\frac{1}{2}\right)}
{\Gamma(3)}\nonumber \\
&-&2 \omega F_D \left(\frac{3}{2},\frac{1}{2},1,\frac{1}{2},\frac{1}{2},2,
\omega\frac{\gamma-r_{\Lambda}^1}{\gamma-\alpha},
\omega\frac{r_{+}-r_{\Lambda}^1}{r_{+}-\alpha},
\omega\frac{r_{\Lambda}^1}{\alpha},
\omega\frac{r_{\Lambda}^2-r_{\Lambda}^1}{r_{\Lambda}^2-\alpha}\right)
\frac{\Gamma\left(\frac{3}{2}\right)\Gamma\left(\frac{1}{2}\right)}
{\Gamma(2)}\nonumber \\
&+& F_D \left(\frac{1}{2},\frac{1}{2},1,\frac{1}{2},\frac{1}{2},1,
\omega\frac{\gamma-r_{\Lambda}^1}{\gamma-\alpha},
\omega\frac{r_{+}-r_{\Lambda}^1}{r_{+}-\alpha},
\omega\frac{r_{\Lambda}^1}{\alpha},
\omega\frac{r_{\Lambda}^2-r_{\Lambda}^1}{r_{\Lambda}^2-\alpha}\right)
\frac{\Gamma\left(\frac{1}{2}\right)\Gamma\left(\frac{1}{2}\right)}
{\Gamma(1)} \nonumber \\
&+&\frac{A^4\sqrt{\omega}}{\sqrt{\frac{\Lambda}{3}}H_{-}}
\nonumber \\
&\times& \Biggl[\omega^2 F_D\left(\frac{5}{2},\frac{1}{2},1,\frac{1}{2},\frac{1}{2},3,
\omega\frac{\gamma-r_{\Lambda}^1}{\gamma-\alpha},
\omega\frac{r_{-}-r_{\Lambda}^1}{r_{-}-\alpha},
\omega\frac{r_{\Lambda}^1}{\alpha},
\omega\frac{r_{\Lambda}^2-r_{\Lambda}^1}{r_{\Lambda}^2-\alpha}\right)
\frac{\Gamma\left(\frac{5}{2}\right)\Gamma\left(\frac{1}{2}\right)}
{\Gamma(3)}\nonumber \\
&-&2 \omega F_D \left(\frac{3}{2},\frac{1}{2},1,\frac{1}{2},\frac{1}{2},2,
\omega\frac{\gamma-r_{\Lambda}^1}{\gamma-\alpha},
\omega\frac{r_{-}-r_{\Lambda}^1}{r_{-}-\alpha},
\omega\frac{r_{\Lambda}^1}{\alpha},
\omega\frac{r_{\Lambda}^2-r_{\Lambda}^1}{r_{\Lambda}^2-\alpha}\right)
\frac{\Gamma\left(\frac{3}{2}\right)\Gamma\left(\frac{1}{2}\right)}
{\Gamma(2)}\nonumber \\
&+& F_D \left(\frac{1}{2},\frac{1}{2},1,\frac{1}{2},\frac{1}{2},1,
\omega\frac{\gamma-r_{\Lambda}^1}{\gamma-\alpha},
\omega\frac{r_{-}-r_{\Lambda}^1}{r_{-}-\alpha},
\omega\frac{r_{\Lambda}^1}{\alpha},
\omega\frac{r_{\Lambda}^2-r_{\Lambda}^1}{r_{\Lambda}^2-\alpha}\right)
\frac{\Gamma\left(\frac{1}{2}\right)\Gamma\left(\frac{1}{2}\right)}
{\Gamma(1)}
\Biggr\}\nonumber \\ 
\label{LambdaPeriastronAdvance}
\end{eqnarray}
where the $H_{+},H^+_{\Lambda},H^-_{\Lambda}$ are given by the 
expressions
\begin{eqnarray}
H_{+}&=&\sqrt{(\alpha-\beta)(\alpha-\gamma)}(\alpha-r_{+})\sqrt{\alpha(
\alpha-r_{\Lambda}^2)} \nonumber \\
H_{\Lambda}^+ &=& (\alpha-r_{\Lambda}^{+})\sqrt{(\alpha-\beta)(\alpha-\gamma)
\alpha(
\alpha-r_{\Lambda}^2)} \nonumber \\
H_{\Lambda}^- &=&(\alpha-r_{\Lambda}^-) \sqrt{(\alpha-\beta)(\alpha-\gamma)
\alpha(
\alpha-r_{\Lambda}^2)} \nonumber \\
\end{eqnarray}
The periastron advance is then given by the expression
\begin{equation}
\Delta\phi^{\rm GTR}_{\rm E}-2 \pi
\label{PeriAdv}
\end{equation}
The phenomenological applications of equations (\ref{LambdaPeriastronAdvance}) and (\ref{PeriAdv}) will be performed systematically in a future publication 
\footnote{The calculation for the corresponding orbital period is 
outlined in appendix \ref{LambdaPeriod}.}.

\subsection{Exact, closed form formula, for the 
 orbital period of equatorial non-circular Kerr geodesics}
\label{XronosIsimerinou}

In this subsection we assume vanishing cosmological constant.
For equatorial orbits $Q$ vanishes and $\theta=\pi/2$.
The relevant differential equation for the calculation of the orbital 
period ${\rm P_E}$ for equatorial non-circular Kerr geodesics is 
\begin{equation}
c\frac{dt}{dr}=\frac{(r^2+a^2)[E(r^2+a^2)-a L]}{\Delta \sqrt{R}}+\frac{a(L-aE)}{\sqrt{R}}
\label{troxisim1}
\end{equation}
Then we want to calculate exactly the following definite integral
\begin{equation}
ct=2\int_{r_P}^{r_A} \frac{(r^2+a^2)[E(r^2+a^2)-aL]}{\Delta \sqrt{R}} dr+
2 \int_{r_P}^{r_A}\frac{a(L-aE)}{\sqrt{R}} dr
\label{TroxiakiIsimeriniP}
\end{equation}
In (\ref{troxisim1}),(\ref{TroxiakiIsimeriniP}) the quartic polynomial $R$ is 
defined in (\ref{LARTH}) with {\boldmath$\Lambda$}$=0,Q=0$.
The idea is to use the appropriate transformation and bring the above 
radial integral into the integral representation of generalized hypergeometric 
function of Appell $F_1$.
We then obtain \footnote{See appendix \ref{IsimTroxPer} for some details 
of the calculation.}
\begin{eqnarray}
ct\equiv c {\rm P_E}&=&\frac{E \beta^2 2 \frac{GM_{BH}}{c^2}}{\sqrt{1-E^2}\sqrt{\alpha-\gamma}\sqrt{\beta-\delta}}
\Biggl[\frac{\Gamma^2 (1/2)}{\Gamma(1)}F_1\left(\frac{1}{2},2,\frac{1}{2},1,\omega,\kappa^2\right) \nonumber \\
&-&\frac{2\omega \gamma}{\beta}\frac{\Gamma(3/2)\Gamma(1/2)}{\Gamma(2)}
F_1\left(\frac{3}{2},2,\frac{1}{2},2,\omega,\kappa^2\right)+\frac{\gamma^2 \omega^2}{\beta^2}\frac{\Gamma(5/2)\Gamma(1/2)}{\Gamma(3)}F_1\left(\frac{5}{2},2,\frac{1}{2},3,\omega,\kappa^2\right)\Biggr]  \nonumber \\
&+&\frac{2 E 2G M_{\rm{BH}}}{c^2} \sqrt{\frac{\omega}{1-E^2}}\frac{\beta}{\sqrt{(\alpha-\beta)(\beta-\delta)}}
\Biggl[\frac{\Gamma^2 (\frac{1}{2})}{\Gamma(1)}F_1\left(\frac{1}{2},1,\frac{1}{2},1,\omega,\kappa^2\right) \nonumber \\
&-&
\frac{\omega \gamma}{\beta} F_1\left(\frac{3}{2},1,\frac{1}{2},2,\omega,\kappa^2\right)
\frac{\Gamma\left(\frac{3}{2}\right)\Gamma\left(\frac{1}{2}\right)}
{\Gamma(2)}\Biggr] \nonumber \\
&+&\frac{4 E 2 G M_{\rm{BH}}}{c^2}\sqrt{\frac{\omega}{1-E^2}}\frac{1}{\sqrt{(\alpha-\beta)(\beta-\delta)}}F\Biggl(\frac{1}{2},\frac{1}{2},1,\kappa^2\Biggr) 
\frac{\Gamma^2(1/2)}{\Gamma(1)} \nonumber \\
&+&\frac{4 E 2 G M_{\rm{BH}}}{c^2} \Biggl[-\frac{\omega^{3/2}A_+}{H_+}
F_1\left(\frac{3}{2},1,\frac{1}{2},2,\kappa^2_+,\mu^2\right)\frac{\Gamma\left(\frac{3}{2}\right)\Gamma\left(\frac{1}{2}\right)}{\Gamma(2)} \nonumber \\
&+&
\frac{\omega^{1/2}A_+}{H_+}F_1\left(\frac{1}{2},1,\frac{1}{2},1,\kappa^2_+,\mu^2\right)\frac{\Gamma^2\left(\frac{1}{2}\right)}{\Gamma(1)}+ \nonumber \\
&-&\frac{\omega^{3/2}A_-}{H_-}
F_1\left(\frac{3}{2},1,\frac{1}{2},2,\kappa^2_-,\mu^2\right)\frac{\Gamma\left(\frac{3}{2}\right)\Gamma\left(\frac{1}{2}\right)}{\Gamma(2)} \nonumber \\
&+&
\frac{\omega^{1/2}A_-}{H_-}F_1\left(\frac{1}{2},1,\frac{1}{2},1,\kappa^2_-,\mu^2\right)\frac{\Gamma^2\left(\frac{1}{2}\right)}{\Gamma(1)}\Biggr] \nonumber \\
&+& a L 2 \frac{G M_{\rm{BH}}}{c^2}\Biggl[-\frac{\omega^{3/2}A_{+}^{\prime}}{H_+}
F_1\left(\frac{3}{2},1,\frac{1}{2},2,\kappa^2_+,\mu^2\right)\frac{\Gamma\left(\frac{3}{2}\right)\Gamma\left(\frac{1}{2}\right)}{\Gamma(2)} \nonumber \\
&+&
\frac{\omega^{1/2}A_{+}^{\prime}}{H_+}F_1\left(\frac{1}{2},1,\frac{1}{2},1,\kappa^2_+,\mu^2\right)\frac{\Gamma^2\left(\frac{1}{2}\right)}{\Gamma(1)}+ \nonumber \\
&-&\frac{\omega^{3/2}A_{-}^{\prime}}{H_-}
F_1\left(\frac{3}{2},1,\frac{1}{2},2,\kappa^2_-,\mu^2\right)\frac{\Gamma\left(\frac{3}{2}\right)\Gamma\left(\frac{1}{2}\right)}{\Gamma(2)} \nonumber \\
&+&
\frac{\omega^{1/2}A_{-}^{\prime}}{H_-}F_1\left(\frac{1}{2},1,\frac{1}{2},1,\kappa^2_-,\mu^2\right)\frac{\Gamma^2\left(\frac{1}{2}\right)}{\Gamma(1)}\Biggr] \nonumber \\
\label{IsimeriniPeriodos}
\end{eqnarray}
where 
\begin{equation}
A_{+}^{\prime}:=\frac{-2 r^{\prime}_+}{r^{\prime}_{-}-r^{\prime}_+},\;A_{-}^{\prime}:=\frac{2 r^{\prime}_-}{r^{\prime}_{-}-r^{\prime}_+}
\end{equation}
and $A_{\pm}$ are given by equation (\ref{Syntelestes1}).
Also $r^{\prime}_{\pm}:=1\pm\sqrt{1-a^2}$, $a$ are the 
dimensionless horizon radii and spin of the black hole respectively.
Equation (\ref{IsimeriniPeriodos}) is the {\em first exact} expression in 
{\em closed analytic form} of the period 
of a test particle in  a non-circular equatorial orbit around a Kerr black 
hole in terms of Appell's generalized hypergeometric function 
$F_1$ and Gau$\ss$'s ordinary hypergeometric function. It 
constitutes a generalization of the Keplerian period for a  
circular equatorial orbit of radius $r$ in the Kerr field \cite{BC} (see also 
Eq.(105) \cite{KraniotisKerr} for $\Lambda=0$)
\begin{equation}
t=2\pi \frac{a\sqrt{\frac{GM_{\rm BH}}{c^2}}\pm r^{3/2}}{c\sqrt{\frac{GM_{\rm BH}}{c^2}}}
\end{equation}
Therefore equation (\ref{IsimeriniPeriodos}) is of paramount importance 
for calculating the orbital and periapsis precession 
periods for non-zero eccentricity equatorial orbits 
of test particles in the vicinity of the horizon $r_+$ of the Kerr black hole
in the strong-field regime of general relativity. 
In addition, it may be of importance  for the phenomenology of 
quasi periodic oscillation (QPO's) frequencies \cite{MAS,Aschenbach} and accretion 
physics.

\subsubsection{Examples of equatorial non-circular orbits with short orbital 
period}

In this subsection we calculate the periapsis advance 
for equatorial non-circular orbits with short orbital periods in 
the asymptotically flat Kerr black hole. In this case
the contribution of the spin of the black hole to the relativistic 
effect of periapsis advance is maximized as compares to the polar orbits. 
Consequently, a precise measurement of the periapsis advance can lead to 
an independent measurement of the spin of the black hole. The exact expression 
for the periapsis advance for the orbit of a test particle in a non-circular 
equatorial motion around a Kerr black hole has been derived in 
\cite{KraniotisLight} and it involves the hypergeometric function $F_1$ of 
Appell \footnote{For the convenience of the reader 
we exhibit this exact formula in appendix \ref{PeriapsisBH}.}. By applying this formula for the choices of initial conditions 
$L=\sqrt{380}\frac{GM_{\rm BH}}{c^2},E=0.99954853$ and for two 
values of the Kerr parameter we obtain the results displayed in Table 
\ref{CuspOrbitEq}. For the calculation of the equatorial orbital period 
$\rm P_E$, eq.(\ref{IsimeriniPeriodos}), for this high-eccentricity 
equatorial orbit we assumed a galactic black hole 
mass $M_{\rm BH}=4.06\times 10^6 M_{\odot}$.
 
\begin{table}
\begin{center}
\begin{tabular}{|l|c|c|}\hline\hline
{\bf Orbital parameters} & {\bf Case 1: a=0.52} & {\bf Case 2: a=0.99616}  \\
$r_P$ &207.748  $\frac{GM_{\rm BH}}{c^2}$  &  207.855$\frac{GM_{\rm BH}}{c^2}$\\
$r_A$ &2005.825  $\frac{GM_{\rm BH}}{c^2}$  & 2005.813 $\frac{GM_{\rm BH}}{c^2}$\\
Periapsis advance:$\Delta\phi^{\rm GTR}_{\rm E}-2\pi$(arcs/rev.) &10082.4 & 9742.34 \\
Periapsis advance ($\frac{\circ}{\rm yr}$) \footnotemark[13]{
Assumes $M_{\rm BH}=4.06\times 10^6 M_{\odot}$}
  &  19.05  &18.4  \\
Orbital Period $\rm {P_E}$ (days) & 53.6 & 53.6\\
\hline \hline
\end{tabular}
\end{center}
\caption{Calculation of orbital data and 
the corresponding relativistic effects for a cusp equatorial  
non-circular stellar orbit around the galactic 
black hole, for two different values of the 
Kerr parameter with $L=\sqrt{380} \frac{G M_{\rm BH}}{c^2},E=0.99954853$.
The exact expression $\Delta\phi^{\rm GTR}_{\rm E}$ is given by Equation (111) in \cite{
KraniotisLight}.}
\label{CuspOrbitEq}
\end{table}
Indeed we observe now the important contribution of the Kerr parameter 
on the relativistic effect of periapsis advance.

A more dramatic example of an equatorial non-circular orbit closer to the 
the event horizon $r_+$ of the galactic rotating black hole is presented in table 
\ref{HorizonOrbitEq}. From the table we see orbital periods in the range 105min-108min depending on the value of Kerr parameter and quite substantial relativistic periapsis advance.  The calculated orbital frequencies are: 0.155mHz, 
0.158mHz for $a=0.52,0.99616$ respectively. For comparison we repeated 
the calculation for a microquasar with black hole mass $M_{\rm BH}=10 
M_{\odot}$ and for $a=0.99616 \frac{GM_{\rm BH}}{c^2}, E=0.9586853, L=\sqrt{13}
\frac{GM_{\rm BH}}{c^2}$. The calculated orbital frequency has the value 
64.2Hz.

\begin{table}
\begin{center}
\begin{tabular}{|l|c|c|}\hline\hline
{\bf Orbital parameters} & {\bf Case 1: a=0.52} & {\bf Case 2: a=0.99616}  \\
$r_P$ & 7.49878 $\frac{GM_{\rm BH}}{c^2}$  & 9.20674 $\frac{GM_{\rm BH}}{c^2}$\\
$r_A$ & 15.1106 $\frac{GM_{\rm BH}}{c^2}$  & 14.1781 $\frac{GM_{\rm BH}}{c^2}$\\
Periapsis advance:$\Delta\phi^{\rm GTR}_{\rm E}-2\pi$(arcs/rev.) &497734 & 309217 \\

Orbital Period ${\rm P_E}$ (s) \footnotemark[14]{
Assumes $M_{\rm BH}=4.060\times 10^6 M_{\odot}$} & $6.462\times 10^3$ & $6.319\times 10^3$\\
Orbital frequency $\rm {\nu_E}$(mHz) & 0.155 &0.158 \\
\hline \hline
\end{tabular}
\end{center}
\caption{Calculation of orbital data and 
the corresponding relativistic effects for a cusp equatorial  
non-circular stellar orbit around the galactic  
black hole for two different values of the 
Kerr parameter with $L=\sqrt{13} \frac{G M_{\rm BH}}{c^2},E=0.9586853$.
The exact expression $\Delta\phi^{\rm GTR}_{\rm E}$ is given by Equation (111)in \cite{
KraniotisLight}.}
\label{HorizonOrbitEq}
\end{table}

\section{Conclusions}

In this work we have investigated the motion of a test particle in Kerr 
spacetime with and without the cosmological constant.

More specifically, we have studied and 
derived for the first time exact solutions of 
the relativistic geodesic differential equations of motion for 
two important classes of possible orbits: a) non-spherical polar Kerr orbits
and b) non-circular equatorial Kerr orbits in the presence of the 
cosmological constant.

In the former case the exact 
orbital solution is given 
in terms of the square of Jacobi's sinus amplitudinous elliptic function.
Furthermore we have derived exact novel expressions for the 
Lense-Thirring effect, 
periapsis advance and orbital period. The resulting solution in closed 
analytic form for the Lense-Thirring gravitomagnetic effect 
 has been expressed in terms of Appell's generalized hypergeometric 
function $F_1$.
The exact formula for the relativistic periapsis advance has been 
expressed elegantly in terms of 
Abel's-Jacobi's amplitude function $am(u,\kappa^2)$ \cite{JACNOTE} with the 
argument $u$
given by Gau$\ss$'s hypergeometric function $F$ 
and the initial conditions for the invariant 
parameters $Q,E$ and the Kerr parameter.
The closed form formulae for the orbit and the periapsis precession have been 
achieved by using the idea of inversion of elliptic integrals 
\cite{NHA2} and properties 
of the modular functions.
The derived solution for the polar orbital period $\rm {P}$ has 
been expressed in terms of Appell's generalized 
hypergeometric 
function $F_1$ of two variables and Gau$\ss$'s hypergeometric function $F$. 
This allowed us to define the LTP (frequency) of polar non-spherical orbits 
around a Kerr black hole.
We also derived the exact solution for the orbital period ${\rm P_E}$ 
for the equatorial non-circular orbit of  a test particle in Kerr spacetime. 
The resulting formula again rests in terms of hypergeometric 
functions of Appell and Gau$\ss$.

Assuming that the galactic centre is a rotating Kerr black hole as is 
indicated by recent observations of the galactic centre, we applied 
our exact solutions for the precise frame dragging and periapsis advance 
of stellar non-spherical polar Kerr orbits near the outer event 
horizon of the supermassive 
black hole. More specifically, we investigated orbits in the central 
milliarcsecond of the galactic centre of Milky Way. Such orbits have short orbital periods 
and therefore if detected they can provide a unique laboratory for measuring 
with high precision the phenomenon of periapsis advance and frame dragging at the strong field 
realm of general relativity. Such angular resolutions will be achieved by 
the GRAVITY experiment using VLBI techniques.
We provided examples with orbital periods in the range 100min-54 days.

We have also calculated the frame dragging and 
periapsis advance effects for the more distant (with respect 
to the outer horizon of the Kerr galactic black hole) S-stars in the central 
arcsecond of Sgr A$^{*}$ for various values of the Kerr parameter and 
the other invariant parameters. 
Since the orbital periods in this case are in the range 14.8 yr-94 yr, 
more precise astrometric data from GRAVITY will allow the probing of 
relativistic effects on a longer ($\sim 10$ year) time scale.

In a similar fashion the exact solution of non-spherical non-polar orbits 
not confined to the equatorial plane has been derived and was expressed 
in terms of Jacobi's sinus amplitudinous elliptic function.

We have also calculated an exact expression for the periastron advance 
of a test particle in a non-circular equatorial orbit around a rotating mass 
in the presence of the cosmological constant (Kerr-de Sitter gravitational 
field). The resulting expression was given in terms of Lauricella's generalized hypergeometric functions $F_D$ of three and four variables.
This solution can be used for investigating in a precise way the 
combined effect of the cosmological constant and the rotation of the central 
mass on the orbits of test particles. 
Our exact solutions of the equations of 
general relativity involve important functions of mathematical analysis 
and the exploration of their topological aspects  constitutes a very interesting topic for future research \cite{KRANIOTIS}. 
 We find very pleasing the productive and exciting interplay 
of observational astronomy of the galactic centre with the exact theoretical 
strong-field regime results presented in this work.

\section*{Acknowledgements} 
This work was partially supported by DOE grant DE-FG03-95-Er-40917. In 
final stages, it was supported by a research fellowship at the Max-Planck-Institut f$\rm \ddot{u}$r Physik in Munich. The author is grateful to the referees for their constructive comments and suggestions that helped improve the presentation of this work. The author wishes to express his gratitude to Professor Dieter 
L$\rm{\ddot u}$st for support and frienship and to Dr J P Bramhall for fixing 
the problem with his arm. He also thanks his wife Rania and son Vassilios for 
support during rehabilitation and Professor Peter McIntyre for kind hospitality.

\appendix

\section{Precise calculation of the orbital period of non-circular 
equatorial geodesics}
\label{IsimTroxPer}

Equation (\ref{TroxiakiIsimeriniP}) can be written
\begin{equation}
c{\rm P_E}=2 \int_{r_P}^{r_A} E(r^2+a^2)\frac{(r^2+a^2)}{\Delta \sqrt{R}}dr
-2 \int_{r_P}^{r_A}\frac{aL(r^2+a^2)}{\Delta \sqrt{R}}dr
-2 \int_{r_P}^{r_A}\frac{a^2 E}{\sqrt{R}}dr+2 \int_{r_P}^{r_A}\frac{aL}{\sqrt{R}}dr
\label{FirstStep}
\end{equation}
Let us outline the calculation of the above definite integral.
The first integral in equation (\ref{FirstStep}) can be written
\begin{eqnarray}
& &\int_{r_P}^{r_A} E(r^2+a^2)\frac{(r^2+a^2)}{\Delta \sqrt{R}}dr \nonumber \\
&=&\int_{r_P}^{r_A}\frac{ E(r^2+a^2)}{\sqrt{R}}dr\left[
1+\frac{2 G M_{\rm BH} \frac{r}{c^2}}{r^2+a^2-2 G M_{\rm BH} \frac{r}{c^2}}\right] \nonumber \\
&=&\int_{r_P}^{r_A}\frac{ E(r^2+a^2)}{\sqrt{R}}dr+
\int_{r_P}^{r_A}\frac{ E }{\sqrt{R}}\frac{2 G M_{\rm BH}r}{c^2}dr+
4 E \left(\frac{GM_{\rm BH}}{c^2}\right)^2
\int_{r_P}^{r_A}\frac{r^2}{\sqrt{R}\Delta}dr \nonumber \\
&=&I_a+4 E \left(\frac{GM_{\rm BH}}{c^2}\right)^2
\int_{r_P}^{r_A}\frac{dr}{\sqrt{R}}-4 E \left(\frac{GM_{\rm BH}}{c^2}\right)^2
\int_{r_P}^{r_A}\frac{a^2-2 G M_{\rm BH} \frac{r}{c^2}}{\Delta \sqrt{R}}dr \nonumber \\
\label{PrwtosOros}
\end{eqnarray}
where 
\begin{equation}
I_a:=\int_{r_P}^{r_A}\frac{ E(r^2+a^2)}{\sqrt{R}}dr+
\int_{r_P}^{r_A}\frac{ E }{\sqrt{R}}\frac{2 G M_{\rm BH}r}{c^2}dr
\end{equation}
Now the term $4 E \left(\frac{GM_{\rm BH}}{c^2}\right)^2
\int_{r_P}^{r_A}\frac{dr}{\sqrt{R}}$ using a dimensionless variable $r^{\prime}$i.e. $r=r^{\prime}\frac{GM_{\rm BH}}{c^2}$ and the transformation 
\begin{equation}
z=\frac{1}{\omega}\frac{r^{\prime}-\alpha_{\mu+1}}{r^{\prime}-\alpha_{\mu+2}}
=\frac{1}{\omega}\frac{r^{\prime}-\beta}{r^{\prime}-\gamma}
\label{Metasximatismos}
\end{equation}
with $\omega=\frac{\alpha-\beta}{\alpha-\gamma}$,
is transformed into the integral representation of Gau$\ss$'s hypergeometric
 function
\begin{eqnarray}
&&4 E \left(\frac{GM_{\rm BH}}{c^2}\right)^2
\int_{r_P}^{r_A}\frac{dr}{\sqrt{R}} \nonumber \\
&=&4 E \frac{GM_{\rm BH}}{c^2}\frac{\sqrt{\omega}}{\sqrt{(1-E^2)(\beta-\delta)
(\alpha-\beta)}}\int_0^1\frac{dz}{\sqrt{z(1-z)(1-\mu^2 z)}} \nonumber \\
&=&4 E \frac{GM_{\rm BH}}{c^2}\frac{\sqrt{\omega}}{\sqrt{(1-E^2)(\beta-\delta)
    (\alpha-\beta)}}F\left(\frac{1}{2},\frac{1}{2},1,\mu^2\right)\pi \nonumber \\
\label{EradiIn65}
\end{eqnarray}
Subsequently let us calculate $I_a$
\begin{equation}
I_a=\int_{r_P}^{r_A}\frac{dr E r^2}{\sqrt{R}}+
\int_{r_P}^{r_A}\frac{dr E a^2}{\sqrt{R}}+
\int_{r_P}^{r_A}\frac{dr E }{\sqrt{R}}\frac{2 G M_{\rm BH}r}{c^2}
\label{Intermediate}
\end{equation}
The second term in equation (\ref{Intermediate}) cancels the third term in 
equation (\ref{FirstStep}).
Now the first term in (\ref{Intermediate}) using the transformation 
(\ref{Metasximatismos}) can be transformed into the integral representation 
of the generalized hypergeometric function $F_1$ of Appell
\begin{eqnarray}
& &\int_{r_P}^{r_A}\frac{dr E r^2}{\sqrt{R}} \nonumber \\
&=&\frac{E G M_{\rm BH} \sqrt{\omega}}{c^2 \sqrt{(1-E^2)(\beta-\delta)(\alpha-\beta)}}
\int_0^1\frac{dz (\gamma \omega z-\beta)^2}{(1-\omega z)^2 \sqrt{
{z(1-z)(1-\mu^2 z)}}} \nonumber \\
&=&\frac{E G M_{\rm BH} \sqrt{\omega}}{c^2 \sqrt{(1-E^2)(\beta-\delta)(\alpha-\beta)}}
\beta^2 \Biggl[\frac{\gamma^2 \omega^2}{\beta^2}F_1\left(
\frac{5}{2},2,\frac{1}{2},3,\omega,\mu^2\right)\frac{\Gamma\left(\frac{5}{2}
\right)
\Gamma\left(\frac{1}{2}\right)}{\Gamma(3)}
\nonumber \\
&+&F_1\left(
\frac{1}{2},2,\frac{1}{2},1,\omega,\mu^2\right)\frac{\Gamma\left(\frac{1}{2}
\right)
\Gamma\left(\frac{1}{2}\right)}{\Gamma(1)} \nonumber \\
&-&\frac{2\omega\gamma}{\beta}F_1\left(
\frac{3}{2},2,\frac{1}{2},2,\omega,\mu^2\right)\frac{\Gamma\left(\frac{3}{2}
\right)
\Gamma\left(\frac{1}{2}\right)}{\Gamma(2)}\Biggr] \nonumber \\
\label{Prerdadin67}
\end{eqnarray}
In a similar fashion the integral $\int_{r_p}^{r_A}
\frac{dr E}{\sqrt{R}}\frac{2 GM_{\rm BH}r}{c^2}$ is calculated to be
\begin{eqnarray}
& &\int_{r_p}^{r_A}
\frac{dr E}{\sqrt{R}}\frac{2 GM_{\rm BH}r}{c^2} \nonumber \\
&=&2 E \frac{GM_{\rm BH}}{c^2} \int_{\beta}^{\alpha} \frac{dr^{\prime} r^{\prime}}
{\sqrt{R^{\prime}}} \nonumber \\
&=&\frac{2 E \frac{GM_{\rm BH}}{c^2}\sqrt{\omega}\beta}
{\sqrt{(1-E^2)(\beta-\delta)(\alpha-\beta)}}\int_0^1\frac{dz(1-\frac{\gamma\omega}{\beta}z)}{(1-\omega z)\sqrt{z(1-z)(1-\mu^2 z)}} \nonumber \\
&=&\frac{2 E \frac{GM_{\rm BH}}{c^2}\sqrt{\omega}\beta}
{\sqrt{(1-E^2)(\beta-\delta)(\alpha-\beta)}}
\Biggl[F_1\left(\frac{1}{2},1,\frac{1}{2},1,\omega,\mu^2\right)
\frac{\Gamma^2\left(\frac{1}{2}\right)}{\Gamma(1)} \nonumber \\
&-&\frac{\gamma\omega}{\beta}F_1\left(\frac{3}{2},1,\frac{1}{2},2,\omega,\mu^2\right)
\frac{\Gamma\left(\frac{3}{2}\right)\Gamma\left(\frac{1}{2}\right)}{\Gamma(2)}
\Biggr] \nonumber \\
&=&\frac{2 E \frac{GM_{\rm BH}}{c^2}\sqrt{\omega}\beta}
{\sqrt{(1-E^2)(\beta-\delta)(\alpha-\beta)}} F_D\left(
\frac{1}{2},1,-1,\frac{1}{2},1,\omega,\frac{\gamma\omega}{\beta},\mu^2\right)
\frac{\Gamma^2\left(\frac{1}{2}\right)}{\Gamma(1)} \nonumber \\
\label{Akakol68}
\end{eqnarray}
where we used the transformation (\ref{Metasximatismos}).

Finally, the last term in equation (\ref{PrwtosOros}) 

\begin{eqnarray}
&-&4 E \left(\frac{GM_{\rm BH}}{c^2}\right)^2
\int_{r_P}^{r_A}\frac{a^2-2 G M_{\rm BH} \frac{r}{c^2}}{\Delta \sqrt{R}}dr \nonumber \\
&=&\frac{4 E  G M_{\rm{BH}}}{c^2} \Biggl[-\frac{\omega^{3/2}A_+}{H_+}
F_1\left(\frac{3}{2},1,\frac{1}{2},2,\kappa^2_+,\mu^2\right)\frac{\Gamma\left(\frac{3}{2}\right)\Gamma\left(\frac{1}{2}\right)}{\Gamma(2)} \nonumber \\
&+&
\frac{\omega^{1/2}A_+}{H_+}F_1\left(\frac{1}{2},1,\frac{1}{2},1,\kappa^2_+,\mu^2\right)\frac{\Gamma^2\left(\frac{1}{2}\right)}{\Gamma(1)}+ \nonumber \\
&-&\frac{\omega^{3/2}A_-}{H_-}
F_1\left(\frac{3}{2},1,\frac{1}{2},2,\kappa^2_-,\mu^2\right)\frac{\Gamma\left(\frac{3}{2}\right)\Gamma\left(\frac{1}{2}\right)}{\Gamma(2)} \nonumber \\
&+&
\frac{\omega^{1/2}A_-}{H_-}F_1\left(\frac{1}{2},1,\frac{1}{2},1,\kappa^2_-,\mu^2\right)\frac{\Gamma^2\left(\frac{1}{2}\right)}{\Gamma(1)}\Biggr] \nonumber \\
\label{Sixtynine}
\end{eqnarray}
where 
$H_{\pm}:=\sqrt{(1-E^2)(\alpha-\beta)}(r^{\prime}_{\pm}-\beta)
\sqrt{\beta-\delta}$.
The moduli (variables) of the hypergeometric function of Appell are 
given by Eq.(\ref{ModuliAppell}).

Likewise we integrate the rest of terms in equation (\ref{FirstStep}) and 
obtain equation (\ref{IsimeriniPeriodos}) for the orbital period 
of a non-circular equatorial orbit.

\subsubsection{Calculation of the angular integrals}
\label{AngularOloklirosi}

In this subsection we present some details of the computation 
that yields the polar orbital period (\ref{PolarOrbitalPeriod}).

The polar orbital period is given by the following definite integral 

\begin{equation}
c{\rm P}=2 \int_{r_P}^{r_A}\frac{dr (r^2+a^2)}{\Delta \sqrt{R}} P_r-
\frac{a^2 E}{\sqrt{Q}}\int_0^\Psi\frac{d\Psi \cos^2 \Psi}{\sqrt{
1-\kappa^{\prime 2}\sin^2 \Psi}}
\end{equation}
where $P_r=E(r^2+a^2)$.
The radial integral is calculated in a similar manner as in the previous 
appendix.
Below we give the details of the angular integration.

Using the latitude variable $\Psi$ the angular integral 
becomes
\begin{equation}
\frac{-a^2 E}{\sqrt{Q}}\int_0^{\varphi}\frac{d\Psi \cos^2(\Psi)}{
\sqrt{1-\kappa^{\prime 2}\sin^2(\Psi)}}=\frac{-a^2 E}{\sqrt{Q}}\int_0^{\varphi}\frac{d\Psi (1-\sin^2(\Psi))}{
\sqrt{1-\kappa^{\prime 2}\sin^2(\Psi)}}
\end{equation}
Now the first angular integral using the variable $x=\sin\Psi$ becomes
\begin{equation}
\frac{-a^2 E}{\sqrt{Q}}\int_0^{\varphi}\frac{d\Psi }{
\sqrt{1-\kappa^{\prime 2}\sin^2(\Psi)}}=\frac{-a^2 E}{\sqrt{Q}}
\int_0^{\sin\varphi}\frac{dx}{\sqrt{(1-x^2)(1-\kappa^{\prime 2} x^2)}}
\end{equation}

Now using a new variable $z$ defined by $\frac{1}{\sin^2 \varphi}x^2=z$
our angular integral becomes
\begin{eqnarray}
& &\frac{-a^2 E}{\sqrt{Q}}\int_0^{\varphi}\frac{d\Psi }{
\sqrt{1-\kappa^{\prime 2}\sin^2(\Psi)}}=\frac{-a^2 E}{\sqrt{Q}}\frac{\sin\varphi}{2}
\int_0^1\frac{dz}{\sqrt{z(1-\sin^2 \varphi\; z)(1-\kappa^{\prime 2}\sin^2 \varphi\; z)}} \nonumber \\
&=&\frac{-a^2 E}{\sqrt{Q}}\frac{\sin\varphi}{2} F_1\left(\frac{1}{2},\frac{1}{2},\frac{1}{2},\frac{3}{2},\sin^2 \varphi,\kappa^{\prime 2}\sin^2 \varphi\right) 2=\frac{-a^2 E}{\sqrt{Q}}
\sin\varphi\; F_1\left(\frac{1}{2},\frac{1}{2},\frac{1}{2},\frac{3}{2},\sin^2 \varphi,\kappa^{\prime 2}\sin^2 \varphi\right)  \nonumber \\
\label{goniaolo}
\end{eqnarray}
This result is used to get (\ref{PERIASTRONSHIFTP1}).
Likewise the second angular integral is calculated elegantly in terms 
of generalized hypergeometric functions of Appell
\begin{eqnarray}
&&\int_0^{\varphi}\frac{d\Psi (-\sin^2 \Psi)}{\sqrt{1-\kappa^{\prime 2}\sin^2(\Psi)}} \nonumber \\
&=&-\frac{\sin^3 \varphi}{2} F_1\left(\frac{3}{2},\frac{1}{2},\frac{1}{2},\frac{5}{2},\sin^2 \varphi,\kappa^{\prime 2}\sin^2 \varphi\right)\frac{2}{3} \nonumber \\
&= &\frac{1}{\kappa^{\prime 2}}\Biggl[\sin\varphi
\;F_1\left(\frac{1}{2},\frac{1}{2},-\frac{1}{2},\frac{3}{2},
\sin^2 \varphi,\kappa^{\prime 2}\sin^2 \varphi\right)-
\sin\varphi \;F_1\left(\frac{1}{2},\frac{1}{2},\frac{1}{2},\frac{3}{2},
\sin^2 \varphi,\kappa^{\prime 2}\sin^2 \varphi\right)\Biggr] \nonumber \\
\label{olokgwniako}
\end{eqnarray}
where in the calculation the following values of the 
Gamma function have been 
used: $\Gamma\left(\frac{1}{2}\right)=\sqrt{\pi},
\Gamma\left(\frac{3}{2}\right)=\sqrt{\pi}/2,\Gamma\left(\frac{5}{2}\right)=
\frac{3 \sqrt{\pi}}{4}, \Gamma(1)=1$.
Use of radial integrations such as (\ref{EradiIn65}),(\ref{Prerdadin67}), 
(\ref{Akakol68}),(\ref{Sixtynine}) and angular 
integrations (\ref{goniaolo}),(\ref{olokgwniako}) yields (\ref{PolarOrbitalPeriod}).

\section{System of differential equations of Lauricella's multivariable hypergeometric 
function $F_D$}
\label{FDLauricella}

In this subsection we give some properties of the $F_D$ function introduced 
in (\ref{HyperLauricella}).
The fourth Lauricella function of $m$-variables is given by \cite{LAURICELLA}
\begin{eqnarray}
F_D(\alpha,{\bf \beta},\gamma;{\bf z})&=&
\sum_{n_1,n_2,\cdots,n_m=0}^{\infty}\frac{(\alpha)_{n_1+\cdots+n_m}
(\beta_1)_{n_1}\cdots(\beta_m)_{n_m}}{(\gamma)_{n_1+\cdots+n_m}(1)_{n_1}\cdots(1)_{n_m}}z_1^{n_1}\cdots z_m^{n_m} \nonumber \\
&=&\sum_{n_1,n_2,\cdots,n_m=0}^{\infty}
\frac{\sum(\alpha,n_1+\cdots n_m)(\beta_1,n_1)\cdots(\beta_m,n_m)z_1^{n_1}\cdots
z_m^{n_m}}{(\gamma,n_1+\cdots +n_m)n_1!\cdots n_m!} \nonumber
\end{eqnarray}
where 
\begin{eqnarray}
{\bf z}&=&(z_1,\cdots,z_m) \nonumber \\
{\bf {\beta}}&=&(\beta_1,\cdots,\beta_m)
\end{eqnarray}
The Pochhammer symbol $(\alpha)_m=(\alpha,m)$ is defined by
$$(\alpha)_m=\frac{\Gamma(\alpha+m)}{\Gamma(\alpha)}=\Bigg \{ \begin{array}{ll}
1,& {\rm if}\;\; m=0, \\
\alpha(\alpha+1)\cdots (\alpha+m-1), & {\rm if}\;\; m=1,2,3,\cdots
\end{array}
$$

It satisfies the following system of differential equations
\begin{eqnarray}
&\cdots&+z_n(1-z_i)\frac{\partial^2 F_D}{\partial z_i \partial z_n}+
[\gamma-(\alpha+\beta_i+1)z_i]\frac{\partial F_D}{\partial z_i}- \nonumber \\
&-& \beta_i z_1\frac{\partial F_D}{\partial z_1}-
\cdots \beta_i z_{i-1}\frac{\partial F_D}{\partial z_{i-1}}-
\beta_i z_{i+1}\frac{\partial F_D}{\partial z_{i+1}}-\cdots \nonumber \\
&&\cdots-\beta_i z_n\frac{\partial F_D}{\partial z_n}-\alpha\beta_iF_D=0
\;\;\;(i=1,2,\cdots,n) \nonumber \\
\end{eqnarray}

The series admits the following integral representation
\begin{equation}
F_D(\alpha,{\bf{\beta}},\gamma;{\bf z})=
\frac{\Gamma(\gamma)}{\Gamma(\alpha)\Gamma(\gamma-\alpha)}
\int_0^1 t^{\alpha-1}(1-t)^{\gamma-\alpha-1}(1-z_1 t)^{-\beta_1}\cdots
(1-z_m t)^{-\beta_m } dt
\end{equation}
which is valid for ${\rm Re}(\alpha)>0, {\rm Re}(\gamma-\alpha)>0$. It converges absolutely inside the $m$-dimensional cuboid:
\begin{equation}
|z_j|<1,\;\;\;\;\;(j=1,\cdots,m)
\end{equation}

\section{Exact orbital solution for non-spherical non-polar Kerr geodesics} 
\label{nosphnp}
\subsection{Contribution to the change in azimuthal angle from 
radial integration for timelike orbits with $L\not =0$}

Using similar techniques as in section \ref{MSPHAIRAPOL} we can 
calculate the contribution to the change of azimuthal angle from the 
radial integration for timelike non-spherical orbits with $L\not =0$.
The corresponding 
radial integral is $\Delta\phi^{r}:=2 \int_{r_P}^{r_A}\frac{-a^2 L+
2 a E\frac{GM_{\rm BH}r}{c^2}}{\Delta\sqrt{R}}dr$ where the quartic 
polynomial $R$ is defined in (\ref{LARTH}) for {\boldmath$\Lambda$}$=0$. 
The above radial integral can be calculated exactly in terms of Appell's 
first generalized hypergeometric function $F_1$ of two variables
\begin{eqnarray}
\Delta\phi^{r}&=&2\Bigl[-\frac{\omega^{3/2}A_+^{NP}}{H_+}
F_1\left(\frac{3}{2},1,\frac{1}{2},2,\kappa^2_+,\mu^2\right)\frac{\Gamma\left(\frac{3}{2}\right)\Gamma\left(\frac{1}{2}\right)}{\Gamma(2)}\nonumber \\
&+&\frac{\omega^{1/2}A_+^{NP}}{H_+}F_1\left(\frac{1}{2},1,\frac{1}{2},1,\kappa^2_+,\mu^2\right)\frac{\Gamma^2\left(\frac{1}{2}\right)}{\Gamma(1)}+ \nonumber \\
&-&\frac{\omega^{3/2}A_-^{NP}}{H_-}
F_1\left(\frac{3}{2},1,\frac{1}{2},2,\kappa^2_-,\mu^2\right)\frac{\Gamma\left(\frac{3}{2}\right)\Gamma\left(\frac{1}{2}\right)}{\Gamma(2)} \nonumber \\
&+&
\frac{\omega^{1/2}A_-^{NP}}{H_-}F_1\left(\frac{1}{2},1,\frac{1}{2},1,\kappa^2_-,\mu^2\right)\frac{\Gamma^2\left(\frac{1}{2}\right)}{\Gamma(1)}\Bigr] \nonumber \\
\label{TimelikeNPolar}
\end{eqnarray}
where now the coefficients $A_{\pm}^{NP}$ are given by the expressions
\begin{equation}
A_{\pm}^{NP}:=-\frac{{\mp}a^2 L\pm 2 a E \frac{GM_{\rm BH}}{c^2}r_{\pm}}{r_{-}-r_{+}}
\end{equation}

\subsubsection{Orbital solution for timelike non-spherical orbits 
with $L\not =0$}
\label{PANSNPNE}

In this subsection and assuming a vanishing cosmological 
constant,  we shall derive the exact solution for a 
non-spherical bound timelike orbit with non-vanishing parameters 
$L$ and $Q$, i.e a non-spherical non-polar and non-equatorial orbit.
The relevant differential equation is
\begin{equation}
\int^r\frac{dr}{\sqrt{R}}=\pm \int^{\theta} \frac{d\theta}{\sqrt{\Theta}}
\label{PANPTM}
\end{equation}
where now the polynomials $R(r)$ and $\Theta(\theta)$ are defined 
in (\ref{LARTH}) for {\boldmath$\Lambda$}$=0$. Using similar steps as in 
section \ref{PeriastronPrecessionPT} we obtain the exact orbital solution in 
terms of the square of Jacobi's sinus amplitudinous function
\begin{equation} 
r=\frac{\beta-\gamma \frac{\alpha-\beta}{\alpha-\gamma}
{\rm sn^2}\left(\frac{\sqrt{1-E^2}\sqrt{(\alpha-\gamma)(\beta-\delta)}}{2}
\int\frac{d\theta}{\sqrt{\Theta}},\kappa^2\right)}{
1-\frac{\alpha-\beta}{\alpha-\gamma}
{\rm sn^2}\left(\frac{\sqrt{1-E^2}\sqrt{(\alpha-\gamma)(\beta-\delta)}}{2}
\int\frac{d\theta}{\sqrt{\Theta}},\kappa^2\right)}\frac{GM_{\rm BH}}{c^2}
\end{equation}
while the angular integration satisfies the equation
\begin{eqnarray}
&&\Biggl[2\times \frac{\sqrt{-a^2(-1+E^2)+L^2+Q+\sqrt{a^4 (-1+E^2)^2-2 a^2 
(-1+E^2)(L^2-Q)+(L^2+Q)^2}}}{2 \sqrt{2}} \nonumber \\
&\times& \frac{4}{\sqrt{1-E^2}}\frac{1}{\sqrt{\alpha-\gamma}}\frac{1}{
\sqrt{\beta-\delta}}\frac{\pi}{2}F\left(\frac{1}{2},\frac{1}{2},1,\kappa^2\right)\Biggr]
\end{eqnarray}
where the modulus of Gau$\ss$'s hypergeometric function is given in 
terms of the roots of the quartic polynomial $R(r)$ by the expression
\begin{equation}
\kappa^2=\frac{\alpha-\beta}{\alpha-\gamma}\frac{\delta-\gamma}{\delta-\beta
} 
\end{equation}
Solving for the angular variable $\Psi$ we obtain the exact expression
\begin{equation}
\frac{\frac{a^2(1-E^2)}{4} \sin^2 \Psi}{e_2-e_3}=
{\rm sn^2}\left[2\sqrt{e_1-e_3}\frac{4}{\sqrt{1-E^2}}\frac{
1}{\sqrt{\alpha-\gamma}}\frac{1}{
\sqrt{\beta-\delta}}\frac{\pi}{2}F\left(\frac{1}{2},\frac{1}{2},1,\kappa^2\right),\kappa^{2\prime\prime}\right]
\label{NPNEVARPSI}
\end{equation}
where the three roots $e_i$ and the modulus $\kappa^{2\prime\prime}$ 
are given by equations (82),(83) in \cite{KraniotisKerr}.

For $L=0$, i.e. the case of polar orbits equation (\ref{NPNEVARPSI}) 
reduces to equation (\ref{PERIASTRONSHIFTP}).

\section{Orbital period for equatorial non-circular Kerr geodesics in 
the presence of the cosmological constant}
\label{LambdaPeriod}

The relevant differential equation is
\begin{equation}
c\frac{dt}{dr}=\frac{1}{\sqrt{R}}\frac{\Xi^2 (r^2+a^2)\left[
(r^2+a^2)E-aL\right]}{\Delta_r}-\frac{1}{\sqrt{R}}a\Xi^2(aE-L)
\end{equation}
where the sextic polynomial is defined in (\ref{LARTH}) with $Q=0$. Then the orbital period $P_E^{\Lambda}$is given by the following integral
\begin{equation}
cP_E^{\Lambda}=2\int_{r_P}^{r_A}(-)\frac{1}{\sqrt{R}}a\Xi^2(aE-L)dr+2
\int_{r_P}^{r_A}\frac{1}{\sqrt{R}}\frac{\Xi^2 (r^2+a^2)\left[
(r^2+a^2)E-aL\right]}{\Delta_r}dr
\end{equation}
The first integral is given by the following exact expression
\begin{eqnarray}
\frac{a\Xi^2 (a E-L)}{\sqrt{\frac{\Lambda}{3}}H}\sqrt{\omega}\Biggl[
-F_D\left(\frac{1}{2},\frac{1}{2},\frac{1}{2},\frac{1}{2},1,\kappa^2,
\lambda^2,\mu^2 \right)\frac{\Gamma(1/2)^2}{\Gamma(1)} \nonumber \\
+\omega F_D\left(\frac{3}{2},\frac{1}{2},\frac{1}{2},\frac{1}{2},2,\kappa^2,\lambda^2,\mu^2\right)\frac{\Gamma(\frac{3}{2})\Gamma(\frac{1}{2})}{\Gamma(2)}
\Biggr] \nonumber \\
\end{eqnarray} 
where $H,\omega$ and the arguments of the generalized hypergeometric function 
of Lauricella are given by equations (\ref{CAPH}),(\ref{OMEGAL}),
(\ref{ARGULAURI}) respectively.
Now we calculate exactly the second term in the integration.
First we have the integral 
\begin{equation}
I_2:=2\int_{r_P}^{r_A}\frac{-aL\Xi^2(r^2+a^2)}{\sqrt{R}\Delta_r}dr
\end{equation}
Using partial fractions 
\begin{equation}
\frac{-aL\Xi^2 (r^2+a^2)}{\Delta_r}=
\frac{A^{1\prime}}{r-r_{\Lambda}^+}+\frac{A^{2\prime}}{r-r_{\Lambda}^-}+
\frac{A^{3\prime}}{r-r_+}+\frac{A^{4\prime}}{r-r_-}
\end{equation}
the coefficients $A^{i\prime},i=1,\cdots,4$ are given by 
\begin{eqnarray}
A^{1\prime}&=&\frac{3 (a^3L\Xi^2+aL\alpha_1^2\Xi^2)}
{(\alpha_1-\beta_1)(\alpha_1-\gamma_1)(\alpha_1-\delta_1)\Lambda} \nonumber \\
A^{2\prime}&=&-\frac{3 (a^3L\Xi^2+aL\delta_1^2\Xi^2)}
{(\alpha_1-\delta_1)(-\beta_1+\delta_1)(-\gamma_1+\delta_1)\Lambda} \nonumber \\
A^{3\prime}&=&-\frac{3 (a^3L\Xi^2+aL\beta_1^2\Xi^2)}
{(-\alpha_1+\beta_1)(-\beta_1+\gamma_1)(\beta_1-\delta_1)\Lambda} \nonumber \\
A^{4\prime}&=&\frac{3 (a^3L\Xi^2+aL\gamma_1^2\Xi^2)}
{(\alpha_1-\gamma_1)(\beta_1-\gamma_1)(\gamma_1-\delta_1)\Lambda} \nonumber \\
\end{eqnarray}
Thus the integral $I_2$ in closed analytic form will be given by generalized 
Lauricella hypergeometric functions $F_D$ of four variables similar to those 
in eq.(\ref{FourRminus}) and (\ref{LambdaPeriastronAdvance}).

In a similar fashion, the partial fraction expansion coefficients 
$A^{i \prime\prime}$ 
for the third integral
\begin{equation}
I_3:=2 \int_{r_P}^{r_A}\frac{\Xi^2 a^2 (r^2+a^2)E}{\sqrt{R}\Delta_r}dr\end{equation}
are given by the expressions
\begin{eqnarray}
A^{1\prime\prime}&=&-\frac{3 (a^4 E \Xi^2+a^2 E \alpha_1^2 \Xi^2}
{(\alpha_1-\beta_1)(\alpha_1-\gamma_1)(\alpha_1-\delta_1)\Lambda} \nonumber \\
A^{2\prime\prime}&=&\frac{3 (a^4 E \Xi^2+a^2 E \delta_1^2 \Xi^2}
{(\alpha_1-\delta_1)(-\beta_1+\delta_1)(-\gamma_1+\delta_1)\Lambda} \nonumber \\A^{3\prime\prime}&=&\frac{3(a^4 E \Xi^2+a^2 E \beta_1^2 \Xi^2}
{(-\alpha_1+\beta_1)(-\beta_1+\gamma_1)(\beta_1-\delta_1)\Lambda} \nonumber \\
A^{4\prime\prime}&=&-\frac{3(a^4 E \Xi^2+a^2 E \gamma_1^2 \Xi^2}
{(\alpha_1-\gamma_1)(\beta_1-\gamma_1)(\gamma_1-\delta_1)\Lambda} \nonumber \\
\end{eqnarray}

The fourth relevant integral is 
\begin{eqnarray}
I_4&:=&2 \int_{r_P}^{r_A}\frac{\Xi^2 r^2 (r^2+a^2) E}{\sqrt{R}\Delta_r}
dr \nonumber \\
&=& 2 \int_{r_P}^{r_A}  \frac{3}{\Lambda}\Xi^2 E\left[-1+\frac{\Delta}{\Delta_r}\right]\frac{1}{
\sqrt{R}}dr \nonumber \\
\end{eqnarray}

The first term in $I_4$ is calculated to be 
\begin{eqnarray}
&-&2 \int_{r_P}^{r_A}  \frac{3}{\Lambda}\Xi^2 E \frac{1}{
\sqrt{R}}dr \nonumber \\
&=&\frac{2\frac{3}{\Lambda}\Xi^2 E }{\sqrt{\frac{\Lambda}{3}}H}\sqrt{\omega}\Biggl[
-F_D\left(\frac{1}{2},\frac{1}{2},\frac{1}{2},\frac{1}{2},1,\kappa^2,
\lambda^2,\mu^2 \right)\frac{\Gamma(1/2)^2}{\Gamma(1)} \nonumber \\
&+&\omega F_D\left(\frac{3}{2},\frac{1}{2},\frac{1}{2},\frac{1}{2},2,\kappa^2,\lambda^2,\mu^2\right)\frac{\Gamma(\frac{3}{2})\Gamma(\frac{1}{2})}{\Gamma(2)}
\Biggr] \nonumber \\
\end{eqnarray}

The partial fraction coefficients for the second term in $I_4$ are
\begin{eqnarray}
A^{1\prime\prime\prime}&=&-\frac{9 (a^2 c^2 E \Xi^2-2 E G M_{\rm BH} \alpha_1 \Xi^2+
c^2 E \alpha_1^2 \Xi^2)}{c^2 (\alpha_1-\beta_1)(\alpha_1-\gamma_1)(\alpha_1-\delta_1)\Lambda^2} \nonumber \\
A^{2\prime\prime\prime}&=&\frac{9 (a^2 c^2 E \Xi^2-2 E G M_{\rm BH} \delta_1 \Xi^2+
+c^2 E \delta_1^2\Xi^2)}
{c^2 (\alpha_1-\delta_1)(-\beta_1+\delta_1)(-\gamma_1+\delta_1)\Lambda^2} \nonumber \\
A^{3\prime\prime\prime}&=&\frac{9(a^2 c^2 E \Xi^2-2 E G M_{\rm BH} \beta_1 \Xi^2+
+c^2 E \beta_1^2\Xi^2)}
{c^2 (-\alpha_1+\beta_1)(-\beta_1+\gamma_1)(\beta_1-\delta_1)\Lambda^2} \nonumber \\
A^{4\prime\prime\prime}&=&-\frac{9 (a^2 c^2 E \Xi^2-2 E G M_{\rm BH} \gamma_1 \Xi^2+
c^2 E \gamma_1^2 \Xi^2)}
{c^2 (\alpha_1-\gamma_1)(\beta_1-\gamma_1)(\gamma_1-\delta_1)\Lambda^2} \nonumber \\
\end{eqnarray}

\section{Periapsis advance for the orbits of S-stars}
\label{PeriapsisBH}

In this appendix in order to gain an appreciation of the magnitude of 
the periapsis advance for the S-stars we perform a precise calculation 
of the effect a) using the exact formula for equatorial non-circular orbits 
produced in \cite{KraniotisLight} and b) using the exact, closed form formula, 
equation (\ref{PERIASTRONSHIFTP}), for non spherical polar orbits derived in 
this work. 
In the former case, we assume that the 
angular momentum axis of the orbit is co-aligned 
with the spin axis of the black hole. In this case the effect of the 
spin of the black hole on the periastron precession is maximized.

For the convenience of the reader we exhibit  below the exact 
formula derived in \cite{KraniotisLight}, which is used in the first 
part of the current 
calculation
\begin{eqnarray}
\Delta\phi^{\rm GTR}_{\rm E}-2\pi&=&\frac{2}{\sqrt{u_1^{\prime}-u_3^{\prime}}}
\frac{1}{\sqrt{\frac{\alpha_S (L-a E)^2}{\left(\frac{GM}{c^2}\right)^3}}}
\Biggl\{\frac{A_{+}}{\frac{GM r_{+}}{c^2 a^2}-u_3^{\prime}}
F_1\left(\frac{1}{2},1,\frac{1}{2},1,\frac{u_2^{\prime}-u_3^{\prime}}
{\frac{GM r_{+}}{c^2 a^2}-u_3^{\prime}},\frac{u_2^{\prime}-u_3^{\prime}}{
u_1^{\prime}-u_3^{\prime}}\right)\pi
\nonumber \\
&+&\frac{A_{-}}{\frac{GM r_{-}}{c^2 a^2}-u_3^{\prime}}
F_1\left(\frac{1}{2},1,\frac{1}{2},1,\frac{u_2^{\prime}-u_3^{\prime}}
{\frac{GM r_{-}}{c^2 a^2}-u_3^{\prime}},\frac{u_2^{\prime}-u_3^{\prime}}{
u_1^{\prime}-u_3^{\prime}}\right)\pi\Biggr\}-2\pi
\nonumber \\
\label{PeriheKraniotisAppell}
\end{eqnarray}
The derivation of the above formula as well 
as the definition of the coefficients $A_{\pm}$ and the cubic polynomial 
whose roots are $u_i^{\prime}$ can be found in \cite{KraniotisLight}, 
pages 4412-4413.

Our free parameters are the Kerr parameter $a$ and the constants of 
integration 
$L,E$.
For a fixed initial choice of values for $L,E$ and for different values of 
the Kerr parameter 
which include those supported from recent observations we 
calculated precisely the resulting periapsis advance, using the 
exact equations 
(\ref{PeriheKraniotisAppell}).
The results are displayed in tables \ref{EINSTEIN1SNPEQ},\ref{EINSTEIN2SNPEQ1},
\ref{PeriapsisHSBH}.
For comparison we note that for Mercury the perihelion advance for 
a particular choice of initial conditions and neglecting the rotation of
the Sun
(see table 1 in \cite{Mercury}), is: 0.103517 $\frac{\rm arcs}{\rm revolution}$.
Thus we conclude that the theoretically calculated values for 
the periastron advance for the S-stars, using 
(\ref{PeriheKraniotisAppell}), displayed 
in tables \ref{EINSTEIN1SNPEQ},\ref{EINSTEIN2SNPEQ1}, are 521-7003 times bigger than 
that of Mercury after one complete revolution. 
From the results of tables  \ref{EINSTEIN1SNPEQ},\ref{EINSTEIN2SNPEQ1} 
we can see the effect of the rotation 
of the black hole. The effect is small and is of the order of a 
few arcseconds per 
revolution, maximized for the stars S2 and S14. From tables 
\ref{EINSTEIN1SNPEQ} and \ref{EINSTEIN2SNPEQ1} we see that S2 and S14 
come closer to the black hole than any other S-star at periastron 
passage. The total relativistic 
periapsis advance decreases by increasing $a$ for 
fixed values of $L,E$.
We repeated the analysis for different values of $L,E$. The results 
are presented in tables \ref{EINSTEIN3SNPEQ3} and 
\ref{EINSTEIN3SNPEQ4}. Our results for the periapsis and apoapsis 
distances are in agreement with the data in \cite{SINFONI}.

Let us compare now the results for the periapsis advance in Tables \ref{EINSTEIN1SNPEQ},\ref{EINSTEIN2SNPEQ1} with the predictions using the exact 
formula, equation  (\ref{PERIASTRONSHIFTP}).
Indeed for the star 
S2, we calculated: $\Delta\Psi^{\rm GTR}=683.916\frac{\rm arcs}{\rm rev},683.894\frac{\rm arcs}{\rm rev}$ respectively for $a=0.52,0.9939$ for the values of $Q,E$ in Tables 
1,2. For S14 we derived the periapsis advance $\Delta\Psi^{\rm GTR}=731.855\frac{\rm arcs}{\rm rev},731.83\frac{\rm arcs}{\rm rev}$ respectively for $a=0.52,0.9939$ for the values of $Q,E$ in Tables 
1,2. This result is expected since for polar orbits the contribution of 
the spin of the black hole is minimal on the effect of periapsis advance.
Our results of the calculation of periapsis advance 
using the closed form formula
(\ref{PERIASTRONSHIFTP}), for the S-stars are displayed in Table \ref{PeriapsisPclosedform}. The predictions for the periapsis and apoapsis distances are 
the same as those in Tables \ref{EINSTEIN1SNPEQ},\ref{EINSTEIN2SNPEQ1}.

On the other hand, let us compare the results of 
tables  for the periastron advance in Tables \ref{EINSTEIN3SNPEQ3} and 
\ref{EINSTEIN3SNPEQ4} with those obtained using (\ref{PERIASTRONSHIFTP}).
For S2 we derived $\Delta\Psi^{\rm GTR}=738.445\frac{\rm arcs}{\rm rev},
738.419\frac{\rm arcs}{\rm rev}$ for $a=0.52,0.9939$ respectively and for the 
choice of values for the parameters $Q,E$ as those displayed in table 
\ref{EINNPolarNStime3a}.  For S14 we derived $\Delta\Psi^{\rm GTR}=926.551\frac{\rm arcs}{\rm rev},
926.511\frac{\rm arcs}{\rm rev}$ for $a=0.52,0.9939$ respectively 
and for the choice of invariant parameters $Q,E$ for S14 as those exhibited in 
table \ref{EINNPolarNStime3a}. \footnote{For non-spherical non-polar orbits 
with orbital inclination $0^{\circ}<i<90^{\circ}$ one expects that the 
resulting periapsis advance has a value between the ones calculated using 
formula (\ref{PERIASTRONSHIFTP})   and  (\ref{PeriheKraniotisAppell}) 
(assuming the 3 orbital configurations, polar, non-polar and 
equatorial have the same eccentricity and 
semi-major axis). A detailed investigation of this matter is however beyond 
the scope of this paper and will be analysed elsewhere.}

The astrometric capability of GRAVITY will also allow deriving very accurate 
orbits for the S-stars. As we saw in the main text the orbital periods in 
this case are in the range of 14.8yr-94yr, and thus more precise astrometric 
data from GRAVITY will allow the probing of general relativistic effects on 
a longer (decade) time scale.

\begin{table}
\begin{center}
\begin{tabular}{|c|c|c|c|c|c|}\hline\hline
{\bf Star}& $L$ & $E$ & $\Delta\phi^{\rm{GTR}}_{\rm E}-2\pi$ & Periapsis & Apoapsis\\\hline
S1 & 267.778473 & 0.999993919 & 54.0874 $\frac{\rm arcs}{\rm revolution}$ & $3.17\times 10^{16}$ cm & $6.69 \times 10^{16}$ cm\\
S2 & 75.4539876 & 0.999979485 & 677.643 $\frac{\rm arcs}{\rm revolution}$ &$1.82\times 10^{15}$ cm & $2.74 \times 10^{16}$ cm \\
S8 & 96.1191430 & 0.999992385 & 418.185 $\frac{\rm arcs}{\rm revolution}$& $2.87\times 10^{15}$ cm & $7.58\times 10^{16}$ cm \\
S12 & 103.1599 & 0.999991213 & 363.183 $\frac{\rm arcs}{\rm revolution}$ & $3.35\times 10^{15}$ cm & $6.49\times 10^{16}$ cm\\
S13 & 192.08634 & 0.99998856 & 105.016 $\frac{\rm arcs}{\rm revolution}$ & $1.58\times 10^{16}$ cm & $3.65\times 10^{16}$ cm\\
S14 & 72.9456205 & 0.999988863 & 724.911$\frac{\rm arcs}{\rm revolution}$ & $1.64\times 10^{15}$ cm & $5.22\times 10^{16}$ cm \\
\hline \hline
\end{tabular}
\end{center}
\caption{Periastron precession for the S-stars in the central arcsecond 
of the galactic centre. The Kerr parameter of the galactic black hole is 
$a_{\rm Galactic}=0.52 \frac{G M_{\rm {BH}}}{c^2}$. The values of the parameter $L$ are in units of  $\frac{G M_{\rm {BH}}}{c^2}$.
We assume a central mass $M_{{\rm BH}}=4.06\times 10^6 M_{\odot}$.}
\label{EINSTEIN1SNPEQ}
\end{table}

\begin{table}
\begin{center}
\begin{tabular}{|c|c|c|c|c|c|}\hline\hline
{\bf Star}& $L$ & $E$ & $\Delta\phi^{\rm{GTR}}_{\rm E}-2\pi$ & Periapsis & Apoapsis\\\hline
S1 & 267.778473 & 0.999993919 & 53.9597 $\frac{\rm arcs}{\rm revolution}$ & $3.17\times 10^{16}$ cm & $6.69 \times 10^{16}$ cm \\
S2 & 75.4539876 & 0.999979485 & 671.947 $\frac{\rm arcs}{\rm revolution}$&$1.82\times 10^{15}$ cm & $2.74 \times 10^{16}$ cm \\
S8 & 96.1191430 & 0.999992385 & 415.429 $\frac{\rm arcs}{\rm revolution}$& $2.87\times 10^{15}$ cm & $7.58\times 10^{16}$ cm  \\
S12 & 103.1599 & 0.999991213 & 360.953 $\frac{\rm arcs}{\rm revolution}$ & $3.35\times 10^{15}$ cm & $6.49\times 10^{16}$ cm \\
S13 & 192.08634 & 0.99998856 & 104.670 $\frac{\rm arcs}{\rm revolution}$& $1.58\times 10^{16}$ cm & $3.65\times 10^{16}$ cm\\
S14 & 72.9456205 & 0.999988863 & 718.607$\frac{\rm arcs}{\rm revolution}$& $1.64\times 10^{15}$ cm & $5.22\times 10^{16}$ cm  \\
\hline \hline
\end{tabular}
\end{center}
\caption{Periastron precession for the S-stars in the central arcsecond 
of the galactic centre. The Kerr parameter of the galactic black hole is 
$a_{\rm Galactic}=0.9939 \frac{G M_{\rm {BH}}}{c^2}$. 
The values of the parameter $L$ are in units of  $\frac{G M_{\rm {BH}}}{c^2}$.
We assume a central mass $M_{{\rm BH}}=4.06\times 10^6 M_{\odot}$. }
\label{EINSTEIN2SNPEQ1}
\end{table}

\begin{table}
\begin{center}
\begin{tabular}{|c|c|c|}\hline\hline
{\bf Star}& case 1:$(a=0.9953)\;\Delta\phi^{\rm{GTR}}_{\rm E}-2\pi$ & case 2:$(a=0.99979)\;\Delta\phi^{\rm{GTR}}_{\rm E}-2\pi$ \\\hline
S1 & 53.9593 $\frac{\rm arcs}{\rm revolution}$ & 53.9581 $\frac{\rm arcs}{\rm revolution}$\\
S2 & 671.930 $\frac{\rm arcs}{\rm revolution}$ & 671.876 $\frac{\rm arcs}{\rm revolution}$ \\
S8 & 415.420 $\frac{\rm arcs}{\rm revolution}$& 415.394$\frac{\rm arcs}{\rm revolution}$ \\
S12 & 360.9469 $\frac{\rm arcs}{\rm revolution}$ & 360.926$\frac{\rm arcs}{\rm revolution}$  \\
S13 &  104.669 $\frac{\rm arcs}{\rm revolution}$ & 104.666$\frac{\rm arcs}{\rm revolution}$ \\
S14 & 718.589$\frac{\rm arcs}{\rm revolution}$ & 718.529 $\frac{\rm arcs}{\rm revolution}$\\
\hline \hline
\end{tabular}
\end{center}
\caption{Periastron precession for the S-stars in the central arcsecond 
of the galactic centre, using the exact, closed-form formula, equation 
(\ref{PeriheKraniotisAppell}), for two 
high values for the spin of the black 
hole. The values for the invariant parameters  $L,E$ are 
the same as those in tables: \ref{EINSTEIN1SNPEQ},\ref{EINSTEIN2SNPEQ1}.
We assume a central mass $M_{{\rm BH}}=4.06\times 10^6 M_{\odot}$.}
\label{PeriapsisHSBH}
\end{table}

\begin{table}
\begin{center}
\begin{tabular}{|c|c|c|c|c|c|}\hline\hline
{\bf Star}& $L$ & $E$ & $\Delta\phi^{\rm{GTR}}_{\rm E}-2\pi$ & Periapsis & Apoapsis\\\hline
S2 & 72.6190898 & 0.999979145 & 731.407$\frac{\rm arcs}{\rm revolution}$ & $1.68\times 10^{15}$ cm & $2.71\times 10^{16}$ cm \\
S14 & 64.8441485 & 0.999987653 & 916.661 $\frac{\rm arcs}{\rm revolution}$ & 
$1.29\times 10^{15}$ cm & $4.73 \times 10^{16}$ cm \\
\hline \hline
\end{tabular}
\end{center}
\caption{Periastron precession for the S-stars in the central arcsecond 
of the galactic centre for different values 
of the parameters $L,E$ than those of table \ref{EINSTEIN1SNPEQ}. The Kerr parameter of the galactic black hole is 
$a_{\rm Galactic}=0.52 \frac{G M_{\rm {BH}}}{c^2}$. We assume a central mass $M_{{\rm BH}}=4.06\times 10^6 M_{\odot}$.}
\label{EINSTEIN3SNPEQ3}
\end{table}

\begin{table}
\begin{center}
\begin{tabular}{|c|c|c|c|c|c|}\hline\hline
{\bf Star}& $L$ & $E$ & $\Delta\phi^{\rm{GTR}}_{\rm E}-2\pi$ & Periapsis & Apoapsis\\\hline
S2 & 72.6190898 & 0.999979145 & 725.018$\frac{\rm arcs}{\rm revolution}$ & $1.68\times 10^{15}$ cm & $2.71\times 10^{16}$ cm \\
S14 & 64.8441485 & 0.999987653 & 907.686 $\frac{\rm arcs}{\rm revolution}$ & 
$1.29\times 10^{15}$ cm & $4.73 \times 10^{16}$ cm \\
\hline \hline
\end{tabular}
\end{center}
\caption{Periastron precession for the S-stars in the central arcsecond 
of the galactic centre for different values 
of the parameters $L,E$ than those of table \ref{EINSTEIN1SNPEQ}. The Kerr parameter of the galactic black hole is 
$a_{\rm Galactic}=0.9939 \frac{G M_{\rm {BH}}}{c^2}$. We assume a central mass $M_{{\rm BH}}=4.06\times 10^6 M_{\odot}$.}
\label{EINSTEIN3SNPEQ4}
\end{table}

\begin{table}
\begin{center}
\begin{tabular}{|c|c|c|}\hline\hline
{\bf Star}& case 1:$(a=0.52)\;\Delta\Psi^{\rm{GTR}}$ & case 2:$(a=0.9939)\;\Delta\Psi^{\rm{GTR}}$ \\\hline
S1 & 54.2277 $\frac{\rm arcs}{\rm revolution}$ & 54.2275 $\frac{\rm arcs}{\rm revolution}$\\
S2 & 683.916 $\frac{\rm arcs}{\rm revolution}$ & 683.894 $\frac{\rm arcs}{\rm revolution}$ \\
S8 & 421.218 $\frac{\rm arcs}{\rm revolution}$& 421.209$\frac{\rm arcs}{\rm revolution}$ \\
S12 & 365.636 $\frac{\rm arcs}{\rm revolution}$ & 365.63$\frac{\rm arcs}{\rm revolution}$  \\
S13 &  105.396 $\frac{\rm arcs}{\rm revolution}$ & 105.395$\frac{\rm arcs}{\rm revolution}$ \\
S14 & 731.855$\frac{\rm arcs}{\rm revolution}$ & 731.83 $\frac{\rm arcs}{\rm revolution}$\\
\hline \hline
\end{tabular}
\end{center}
\caption{Periastron precession for the S-stars in the central arcsecond 
of the galactic centre, using the exact, closed-form formula, equation 
(\ref{PERIASTRONSHIFTP}), for two 
different values for the spin of the black 
hole. The values for the invariant parameters  $Q,E$ are 
the same as those in tables: \ref{EINNPolarNStime1},\ref{EINNPolarNStime2}.
We assume a central mass $M_{{\rm BH}}=4.06\times 10^6 M_{\odot}$.}
\label{PeriapsisPclosedform}
\end{table}


\newpage

\begin{thebibliography}{99}



\bibitem{Albert} A. Einstein, {\em Erkl${\rm \ddot a}$rung der 
Perihelbewegung des Merkur aus der allgemeinen Relativit${\rm \ddot a}$tstheorie}, \emph{
Sitzungsberichte der Preussischen Akademie der Wissenschaften},(1915) 831.


\bibitem{Karl} K. Schwarzschild, {\em ${\rm \ddot{U}}$ber das Gravitationfeld 
eines Massenpunktes nach der Einsteinschen Theorie.} Sitzungsberichte der 
K${\rm \ddot{o}}$niglichen Preussischen Akademie der Wissenschaften (Berlin) 
1916,189-196

\bibitem{KERR} R. P. Kerr, {\em Gravitational field of a spinning mass 
as an example of algebraically special metrics}, Phys. Rev. Letters 11, (1963),237

\bibitem{OHANIAN} H. Ohanian \& R. Ruffini, Gravitation \& Spacetime, Norton
and Company, New York, 1994.


\bibitem{REES} M. J. Rees, {\em Black Holes in the Real Universe and 
their prospects as Probes of Relativistic Gravity}, astro-ph/0401365



\bibitem{Mercury} G. V. Kraniotis, S. B. Whitehouse{ \em 
Compact calculation of the perihelion precession of Mercury in 
general relativity, the cosmological constant and Jacobi's inversion 
problem}, Classical and Quantum Gravity ${\bf 20}$, (2003) 4817-4835.

\bibitem{BAHCALL} N.A. Bahcall, \ J.P. Ostriker, S. Perlmutter and P.J.
Steinhardt, Science 284(1999)1481

\bibitem{PERLMUTTER} S. Perlmutter et al. Astrophys.J. 517(1999),565;
A.V. Filippenko et al, Astron.J. 116(1998)1009

\bibitem{DEBERNAR} P. deBernardis et al, Nature 404(2000)955; A.H. Jaffe et
al, Phys.Rev.Lett.86(2000)3475

\bibitem{GVKSBW} G. V. Kraniotis and S. B. Whitehouse, {\em General 
relativity, the cosmological constant and modular forms}, {\rm
Class. Quantum Grav.}{\bf 19} (2002)5073-5100, [arXiv:gr-qc/0105022]






\bibitem{KraniotisKerr} G. V. Kraniotis, {\em Precise relativistic orbits 
in Kerr and Kerr-(anti) de Sitter spacetimes}, {\rm Class. Quantum Grav.}
{\bf 21} (2004) 4743-4769, [arXiv:gr-qc/0405095]

\bibitem{KraniotisLight} G. V. Kraniotis, {\em Frame dragging and 
bending of light in Kerr and Kerr-(anti) de Sitter spacetimes}, {\rm Class. Quantum Grav.} {\bf 22} (2005) 4391-4424, [arXiv:gr-qc/0507056]









\bibitem{Schoedel} R. Sch${\rm \ddot{o}}$del {\em et al} Nature 419,(2002)694-696, K. Gebhardt, Nature 419, (2002) 675-676


\bibitem{GHEZa} A M Ghez et al, ApJ {\bf 586} (2003) L127-L131

\bibitem{GENZEL} R. Genzel et al, {\em Near-Infrared flares from accreting 
gas around the supermassive black hole at the Galactic  Centre}, Nature, 
Vol.425 (2003), 934-937. 


\bibitem{Porquet} B. Aschenbach, N. Grosso, D. Porquet and P. Predehl,{
\em X-ray flares reveal mass and angular momentum of the Galactic Centre 
black hole}, astro-ph/0401589, {\em Astron. Astrophys.} {\bf 417} 2004 pp 
71-78

\bibitem{GRAV} F. Eisenhauer et al, {\em GRAVITY: The AO-Assisted, Two-Object Beam-Combiner Instrument for the VLTI}, astro-ph/0508607


\bibitem{SINFONI} F. Eisenhauer et al, {\em Sinfoni in the galactic centre:
Young stars and infrared flares in the central light-month}, ApJ{\bf 628} 
(2005), 246-259.

\bibitem{Paumard} T. Paumard et al, ApJ{\bf 643}, (2006) 1011-1035

\bibitem{LB} Y. Levin and A M Beloborodov, ApJ{\bf 590}, (2003) L33

\bibitem{MJREIDHAFT} M. J. Reid et al, ApJ{\bf 524} (1999), 816-823.


\bibitem{GEOMRO} F. Eisenhauer et al, {\em A geometric determination of the 
distance to the galactic centre}, ApJ{\bf 597} (2003) L121-L124

\bibitem{REID} M. J. Reid, Annu. Rev. Astron.Astrophys. 31 (1993) 345-72;
J. H. Oort, L. Plaut, Astron.\& Astrophys.{\bf 41} (1975), 71-86

\bibitem{SPIN} J. M. Bardeen, W. M. Press and S. A. Teukolsky, {\em 
Rotating black holes: Locally, non-rotating frames, energy extraction and 
scalar synchrotron radiation}, Astrophys.J.{\bf 178}, (1972) 347-369;
F. Melia,  C. Bromley, S. Liu and C. K. Walker, {\em Measuring the 
black hole spin in ${\rm SgrA^{\star}}$}, Astrophys.J.{\bf 554}, L37-L40 
(2001)


\bibitem{BOPR} R. H. Boyer and T. G. Price, {\em An interpretation of the 
Kerr metric in general relativity}, Proc.Camb.Phil.Soc.(1965),{\bf 61}, 531


\bibitem{LTPr} J. Lense, H. Thirring, {\em $\ddot{U}$ber den Einflu$\ss$ 
der Eigenrotation der Zentralk$\ddot {o}$rper auf die Bewegung der Planeten und Monde nach der Einsteinschen Gravitationstheorie}, Phys.Zeitsch.{\bf 19}, (1918) 156




\bibitem{Aschenbach} B. Aschenbach, 
%{\em Measuring mass and angular momentum 
%of black holes with high-frequency quasi-periodic oscillations},
{\em Astron. Astrophys.} {\bf 425} 2004 pp 1075-1082, astro-ph/0406545

\bibitem{ZdenekA} Z. Stuchl\'ik, P. Slan\'i, G. T$\rm{\ddot o}$r$\rm{\ddot o}$k, M. Abramowicz, Phys.Rev.D 71, 024037 (2005); Z. Stuchl\'ik, P. Slan\'i, G. T$\rm{\ddot o}$r$\rm{\ddot o}$k,Proceedings of Science, VI Microquasar Workshop:
{\em Microquasars and Beyond}, Sept.2006, Como, Italy  














\bibitem{Boyer} R. H. Boyer and R. W. Lindquist, {\em Maximal Analytic 
Extension of the Kerr metric}, Journal of Math.Phys.8 (1967), 265-281.


\bibitem{Zdenek} Z. Stuchlik, M. Calvani, {\em Null Geodesics in Black 
Hole metrics with non-zero Cosmological Constant}, General Relativity and 
Gravitation, {\bf 23} (1991), 507-519; P. Slany and Z. Stuchlik, 
{\rm Class. Quantum Grav.} {\bf 22} (2005) 3623-3651

\bibitem{CARTER} B. Carter, {\em Hamilton-Jacobi and Schrodinger Separable 
Solutions of Einstein's Equations} Commun.Math.Phys. 10, (1968), 280-310;
M. Demianski, Acta Astronomica, Vol.23(1973) 197-231;
S. W. Hawking, C. J. Hunter and M. M. Taylor-Robinson, Phys.Rev.D 59 (1999),
064005


%%%\bibitem{Everitt} 
%%{\em Gyros, Clocks, Interferometres\dots:Testing Relativistic Gravity in Space%%}
%%Lect.Notes in Physics 562. Springer Berlin, (2001) pp52-82. Eds C. L$\rm \ddot%%{a}$mmerzahl, C.W.F. Everitt and F. W. Hehl;
%%R. A. Van Patten and C.W.F. Everitt, {\em Possible Experiment with Two Counter%%-Orbit 
%%Drag-free Satellites to obtain a New Test of Einstein's General Theory 
%%of Relativity and improved measurements in Geodesy}, Phys.Rev.Lett.36 (1976)62%%9-%%632



%%\bibitem{CIU} I. Ciufolini,{\em Measurement of the Lense-Thirring Drag on 
%%High-Altitude, Laser-Ranged Artificial Satellites}, Phys.Rev.Let.{\bf 56} 1986%%,
%%278-281; I. Ciufolini, gr-qc/0209109; I. Ciufolini and E. C. Pavlis, Nature 
%%{\bf 431} (2004), 958-960



\bibitem{NHA2} N. H. Abel, {\em Pr\'{e}cis d'une th\'{e}orie des fonctions 
elliptiques},  Crelle's Journal fuer Math. {\bf 3}, (1828) 236-277;
Crelle's  Journal fuer Math.{\bf 4}, (1829) 309-348






\bibitem{JACNOTE} J. Jacobi, {\em Note sur les fonctions elliptiques}, 
Crelle's Journal f$\rm \ddot{u}$r Math. {\bf 3} 1828, 192-195







\bibitem{carter2} B. Carter, {\em Global Structure of the Kerr family of 
Gravitational fields}, Phys.Rev. {\bf 174}, (1968), p 1559-1571

\bibitem{BC} C T Cunningham and J M Bardeen, {\bf ApJ} 173 (1972) L137


\bibitem{JMR} M J Rees, Nature 333 (1998) 523


\bibitem{LAURICELLA} G. Lauricella, {\em Sulle funzioni ipergeometriche a 
pi\`{u} variabili}, Rend. Circ. Mat. Palermo {\bf 7} (1893), 111-158












\bibitem{APPELL} P. Appell, {\em Sur les fonctions hyperg\'{e}om\'{e}triques 
de deux variables}, Journal de Math\'{e}matiques pures et appliqu\'{e}es 
de Liouville, VIII (1882) p 173-216



\bibitem{MAS} G. T$\rm{\ddot o}$r$\rm{\ddot o}$k, M. Abramowicz, W. 
Klu\'zniak, Z. Stuchl\'ik, astro-ph/0603847

%%\bibitem{AECKART} G. F. Rubilar and A. Eckart, {\em Astron. Astrophys.} {\bf 
%%374} (2001) 95-104; P. C. Fragile and G. J. Mathews, ApJ{\bf 542} (2000) 328-3%%33


\bibitem{StoTsou} E. Stoghianidis and D. Tsoubelis, {\em Polar Orbits in 
the Kerr Space-Time}, General Relativity and Gravitation, Vol.19 (1987),p 1235-1249; E. Stogianidis and D. Tsoubelis, Phys.Lett.A {\bf 116} 1986, 213-215



\bibitem{KUMMER} E. E. Kummer, {\em ${\ddot U}$ber die hypergeometrische 
Reihe $1+\frac{\alpha\beta}{1.\gamma}x+\frac{\alpha(\alpha+1)\beta(\beta+1)}
{1.2.\gamma(\gamma+1)}x^2+ \cdots$} Crelle's Journal.f. Math.{\bf 15} (1836) 
39-172



%%\bibitem{TARTACARDA} G. Cardano 1545, \emph{Artis Magnae sive de regvlis
%%algebraicis.}; G Cardano 1968, English translation:\emph{The Great Art or the %%Rules of
%%%Algebra T R Witmer }MIT Press, Cambridge, MA


\bibitem{KRANIOTIS} G.V. Kraniotis, Work in progress.












 






\end{thebibliography}
\end{document}